\documentclass[useAMS,usenatbib]{mn2e}
\usepackage{graphicx,natbib,times,mathptm,bm, amssymb, bibentry, ifpdf, datetime,hyperref, multirow}
\usepackage[fleqn]{amsmath}
\pdfoutput=1
{\newif\ifnotend
\notendtrue
\def\veclist{ABCDEFGHIJKLMNOPQRSTUVWXYZabcdefghijklmnopqrstuvwxyz.}
\def\top#1#2.{#1}
\def\tail#1#2.{#2.}
\loop\expandafter\xdef\csname v\expandafter\top\veclist\endcsname%
{{\noexpand\bm\expandafter\top\veclist}}
\edef\veclist{\expandafter\tail\veclist}
\if\veclist.\notendfalse\fi\ifnotend\repeat}
\let\boldgrk=\gkvecten
\let\boldgrksc=\gkvecseven
\def\gkthing#1{{\mathchoice%
        {\hbox{{\boldgrk\char#1}}}
        {\hbox{{\boldgrk\char#1}}}
        {\hbox{{\boldgrksc\char#1}}}
        {\hbox{{\boldgrksc\char#1}}}}}

\def\vtheta{\gkthing{18}}

\def\d{{\rm d}}

\def\pr{{\mathop{\hbox{p}}}}

\def\Halpha{{\rm H}\alpha}
\def\Hbeta{{\rm H}\beta}

\newcommand{\appropto}{\mathrel{\vcenter{
  \offinterlineskip\halign{\hfil$##$\cr
    \propto\cr\noalign{\kern2pt}\sim\cr\noalign{\kern-2pt}}}}}

\bibpunct{(}{)}{;}{a}{}{,}

\def\galaxymodel{\beta}
\def\transfermodel{\alpha}

\def\feh{[{\rm Fe}/{\rm H}]}
\def\mh{[{\rm M}/{\rm H}]}
\def\afe{[\alpha / {\rm Fe} ]}

\def\Cmat{\textsf{\textbf{C}}}

\def\extnlin{c^{(1)}}
\def\extnquad{c^{(2)}}

\bibpunct{(}{)}{;}{a}{}{,}

\begin{document}

\bibliographystyle{mn2e}

\title[Marginal likelihoods of distances and extinctions to stars]
{Marginal likelihoods of distances and extinctions to stars:
  computation and compact representation}

\author[Sale and Magorrian]{S.~E.~Sale${}^1$ and J. Magorrian${}^{1,2}$\\
  $^1$Rudolf Peierls Centre for Theoretical Physics, Keble Road, Oxford
  OX1 3NP, UK\\
$^2$Institut d'Astrophysique de Paris, 98bis Boulevard Arago, 75014,
Paris, France}

\date{Received .........., Accepted...........}

\maketitle

\begin{abstract}
  We present a method for obtaining the likelihood function of
  distance and extinction to a star given its photometry.
  The other properties of the star (its mass, age, metallicity and so
  on) are marginalised assuming a simple Galaxy model.
  We demonstrate that the resulting marginalised likelihood function
  can be described faithfully and compactly using a Gaussian mixture
  model.
  For dust mapping applications we strongly advocate using
  monochromatic over bandpass extinctions, and provide tables for
  converting from the former to the latter for different stellar
  types.

\end{abstract}

\section{Introduction}

Our present lack of knowledge of the three-dimensional distribution of
interstellar dust is a significant barrier to building a complete
picture of our Galaxy.
Like the sun, most of the Galaxy's stars lie close to the plane, which
means that their light is subject to significant extinction before it
reaches us.
Therefore any attempt to construct a complete model of the Galaxy's
stellar density distribution must include a three-dimensional model of
the extincting dust distribution.
Dust is also interesting in its own right as a tracer of the densest
parts of the interstellar medium (ISM).
Consequently there is now a growing industry devoted to understanding
and mapping extinction, with a number of authors either presenting
methods for mapping extinction \citep[e.g][]{Majewski_Zasowski.2011,
  Sale_only.2012, Hanson_Bailer-Jones.2014, Green_Schlafly.2014} or
constructing actual maps of extinction in two or three dimensions
\citep[e.g.][]{Marshall_Robin.2006, PlanckCollaboration_Abergel.2014a,
  Lallement_Vergely.2014, Sale_Drew.2014}.

A superficially attractive and straightforward way of producing a 3D
extinction map is by first using a method such
as \cite{Berry_Ivezic.2012} and \cite{Hanson_Bailer-Jones.2014} to
calculate posterior expectations for the distances and extinctions to
large numbers of stars individually, then binning the results
spatially to produce a map.
Unfortunately, this produces maps that are biased in a complicated
manner.
There are three principal sources of bias.
First, almost any catalogue of stars will itself not be an unbiased
sample of the stars in the Galaxy.
Most catalogues are magnitude limited, which biases them towards
less extinguished, and therefore brighter, stars.
Dealing effectively with such selection effects is not trivial and is
the subject of Sale (in prep.).
Second, we expect that extinction along two nearby sightlines should be
correlated: two-dimensional projected dust maps exhibit correlations
on scales ranging from less than 1~pc 
\citep[e.g.][]{DiFrancesco_Sadavoy.2010} up to that of spiral arms.
Third, the posteriors distributions of the distances and extinctions 
to individual stars are frequently extended and exhibit complicated 
forms.
As a result the posterior expectations of distance and extinction will
not transmit the full range of uncertainties nor the complex 
correlations that exist between distance and extinction.

In \cite{Sale_Magorrian.2014} we presented a new method for mapping
extinction from star counts that avoids these problems.
Building on earlier work in \cite{Vergely_FreireFerrero.2001} and
\cite{Sale_only.2012}, we used a simple physical model of Kolmogorov
turbulence to impose spatial correlations on the density map, which
prevents the formation of non-physical `fingers of God' as found in the
maps in \cite{Marshall_Robin.2006} and \cite{Sale_Drew.2014}.
The method avoids the need for spatial binning, producing
(realisations of) extinction maps whose resolution is set naturally by
the available data.

Most of the three-dimensional extinction mapping procedures mentioned
above, including that of \cite{Sale_Magorrian.2014}, share the
requirement that one have some way of calculating the marginal
likelihoods of distances and extinctions to individual stars.
That is, having some observations $\vy$ (photometry and/or
spectroscopy and/or astrometry) of a single star at Galactic 
coordinates $(l,b)$, they need the likelihood 
$\pr(\vy|s,l,b,A,\transfermodel,\galaxymodel)$ of the distance~$s$ and
extinction~$A$ to the star, in which the details of the star's mass, 
age, metallicity and so on have been marginalised out assuming some 
galaxy model~$\galaxymodel$ and set of extinction laws and 
isochrones~$\transfermodel$.
The present paper provides one way of calculating such marginalised
likelihoods.
We begin though by considering the problem of how best to parametrize
the extinction law included in~$\transfermodel$ and how to calculate 
the effects of extinction in a range of popular photometric passbands.
This is the subject of section~\ref{sec:parametrizing}; tables giving
the results of our calculation are available online.
Then in section~\ref{sec:prob} we present a method for calculating the
marginal likelihood\footnote{We note that `marginal likelihood' can 
refer to a likelihood with some \emph{or} all of the parameters of the
model employed marginalised out.
However, this terminology is most frequently used in the case where 
all parameters have been marginalised, in which case the `marginal 
likelihood' is sometimes also called the `evidence' and is employed in 
model selection applications.
Our use of the term is distinct to this case as we only marginalise 
some parameters.} $\pr(\vy|s,l,b,A,R,\transfermodel,\galaxymodel)$ and
constructing compact, accurate fits to its dependence on $(s,A)$.
Section~\ref{sec:summary} sums up.

\section{Parametrizing Extinction}\label{sec:parametrizing}

We start by defining extinction and its relationship to the column of
dust between us and a star.
Much of what we discuss in this section has previously appeared by
various authors, including \cite{Golay_only.1974},
\cite{McCall_only.2004}, \cite{Sale_Drew.2009},
\cite{Stead_Hoare.2009}, \cite{Bailer-Jones_only.2011} and
\cite{Casagrande_VandenBerg.2014}.
None the less, we repeat it here for completeness and clarity.

Historically, extinction has usually been estimated by looking at the
broad-band colours of stars.
If one has a star of known spectral type, then by
comparing, say, the measured $B-V$ colour of the star to its expected
intrinsic colour, one obtains the colour excess
\begin{equation}
  E(B-V)\equiv(B-V)_{\rm measured}-(B-V)_{\rm intrinsic},
\end{equation}
which is a direct estimate of the difference $A_B-A_V$ between the
$B$- and $V$-band extinctions to the star.
Typically \citep[e.g.][]{Cardelli_Clayton.1989, Fitzpatrick_only.2004}
the shape of the extinction law at optical and near--infrared
wavelengths (see Figure~\ref{fig:extn_laws} below) is assumed to
depend on a single parameter $R_V$, defined through
\begin{align}
R_V \equiv \frac{A_V}{A_B-A_V} = \frac{A_V}{E(B-V)}.
\label{eqn:RV}
\end{align}
Estimates of $A_V$, $A_B$ then follow directly from $E(B-V)$ given an
assumed $R_V$.
The procedure for other bands $(X,Y)$ is similar: measure a colour
excess $E(X-Y)$, then use an assumed extinction law to obtain the
broad-band extinctions $A_X$ and $A_Y$.

Such broad band extinctions are less than ideal for mapping dust,
however.
To see this, recall that the extinction in a band $X$ to a distance
$s$ along a single line of sight\footnote{To keep this and subsequent
  expressions readable, we suppress the dependence on the
  line of sight $(l,b)$ in this and subsequent expressions.} is given by
\begin{align}
A_X(s) = -2.5 \log_{10} \left( \frac{\int_0^{\infty} \, d\lambda  F(\lambda) T_X(\lambda) e^{- \int_0^s \,ds' \kappa_\lambda(s') \rho(s')}}{\int_0^{\infty} \,d\lambda F(\lambda) T_X(\lambda)  } \right) , \label{eqn:AX_defn}
\end{align}
where $\rho(s)$ is the density of dust along the line of sight,
$\kappa_\lambda(s)$ is its wavelength-dependent opacity, $F(\lambda)$
is the SED of the observed star and $T_X(\lambda)$ the combination of
the transmission of the filter $X$, the transmission of the atmosphere,
the transmission of the rest of the telescope, the detector efficiency
and a function that characterises how the detector responds to 
incident flux\footnote{ For
  most modern detectors this is a factor proportional to $\lambda$
  since the detector counts incident photons
  \citep{Bessell_only.2005}.}.
It is obvious from this equation that passband-based measurements of
extinction and reddening, such as $A_V$ and $E(B-V)$, depend not only
on the dust column between us and a star (i.e. $\rho$ and $\kappa$),
but also on the star's SED 
\citep[see also e.g][]{McCall_only.2004, Sale_Drew.2009,
Bailer-Jones_only.2011}.
Consequently it is possible to observe two stars of different spectral
types behind the exact same dust column and obtain different
measurements of, e.g., $A_V$ from each.
It is perhaps less immediately obvious from
equation~\eqref{eqn:AX_defn} that the relationship between column
density and the broad-band extinction $A_X$ is not linear, even when
the opacity $\kappa_\lambda$ is independent of position.
Therefore, although quantities such as $E(B-V)$ and $A_V$ are
conveniently close to observation, they mix together the effects of
the dust column and stars' SEDs in a way that is non-trivial to
disentangle: by considering passband-based measurements of extinction
one obscures the true physics of the ISM behind a layer of obfuscating
variables.

An alternative to passband-based measurements of extinction is to
consider monochromatic measurements \citep[e.g.,][]{McCall_only.2004,Sale_Drew.2009,Bailer-Jones_only.2011}.
The monochromatic extinction at wavelength~$\lambda$ is given by
\begin{align}
A_{\lambda}(s) &= -2.5 \log_{10} \left(  e^{- \int_0^s \,ds\, \kappa_\lambda(s') \rho(s')} \right) \\
&= 1.086 \int_0^s \,ds' \kappa(s',\lambda) \rho(s'), \label{eqn:Alambda_defn}
\end{align}
which follows directly from~\eqref{eqn:AX_defn} on adopting a Dirac
delta function for the transmission filter~$T_X(\lambda)$.
It is immediately apparent that this $A_{\lambda}(s)$ does not depend
on the SED of the observed star and that its derivative $\d
A_\lambda/\d s$ is linear in $\kappa_\lambda\rho$.
Therefore monochromatic extinction offers a much more direct view on
the distribution of dust, mediated only by variations in dust opacity.

It might appear that using monochromatic extinctions would be
significantly more complicated than employing band-based measurements.
But, if working within a Bayesian framework, or indeed with any
methodology that employs a forward model, building in a monochromatic
measure of extinction is essentially trivial: one simply requires a
model for how variations in monochromatic extinction will alter
observed apparent magnitudes.
We will develop this model in section~\ref{sec:A_response}.

\subsection{The wavelength dependence of extinction}

\begin{figure}
\includegraphics{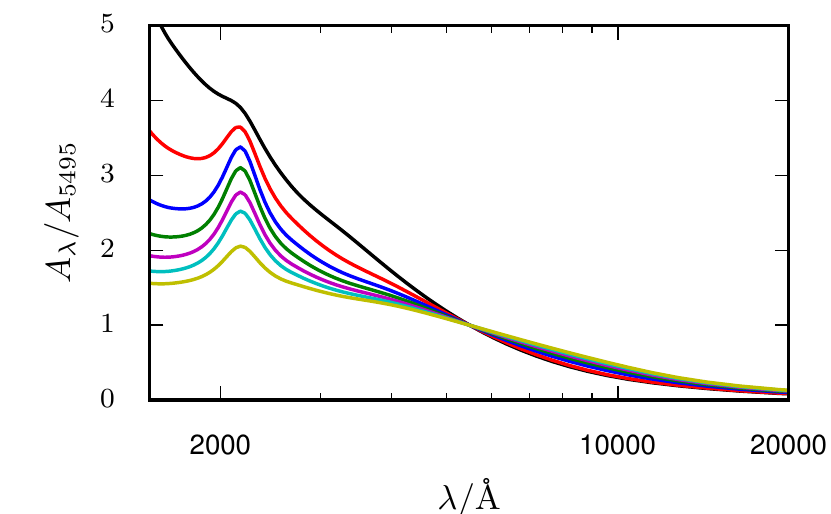}
\caption{\protect\cite{Fitzpatrick_only.2004} extinction laws for
`$R_V=2.1$' (black), `$R_V=2.6$' (red), `$R_V=3.1$' (blue), `$R_V=3.6$' (green), `$R_V=4.1$' (magenta), `$R_V=4.6$' (cyan) and `$R_V=5.1$' (yellow).
All have been normalised to 5495~\AA.
\label{fig:extn_laws}}
\end{figure}

\begin{figure}
\includegraphics{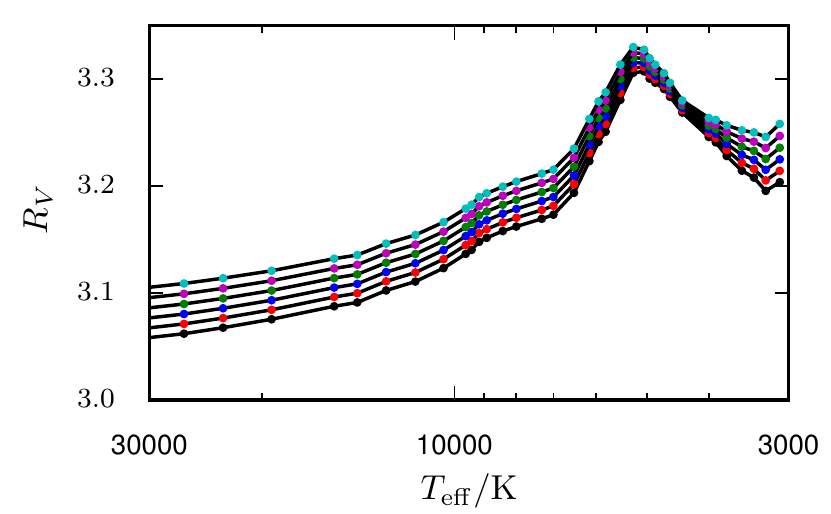}
\caption{The $R_V$s implied by column densities of $R_{5495}=3.056$ dust as a function of effective temperature along the main sequence defined by \protect\cite{Straizys_Kuriliene.1981}. 
Different colours correspond to different extinctions: $A_{4000}=0$ (black), 2 (red), 4 (blue), 6 (green), 8 (magenta) and 10 (cyan).
\label{fig:RV_implied}}
\end{figure}

The normalised form of the opacity ($\kappa$) dependence on
wavelength is typically referred to as the reddening or extinction
law, the shape of which depends on the dust grain size distribution,
with a greater number of larger grains leading to a greyer extinction
law \citep{Weingartner_Draine.2001}.
There exist a number of parametrizations of extinction laws, inferred 
from a range of sightlines within the Galaxy 
\citep[e.g.][]{Cardelli_Clayton.1989, ODonnell_only.1994, 
  Fitzpatrick_only.2004}.
In Fig.~\ref{fig:extn_laws} we show a number of those given by 
\cite{Fitzpatrick_only.2004}.

Typically \citep[e.g.][]{Cardelli_Clayton.1989, Fitzpatrick_only.2004}
the shape of extinction law at optical and near--infrared wavelength
is assumed to be a function of the single parameter~$R_V$ defined in
equation~\eqref{eqn:RV} above.
This is not ideal, as the broadband extinctions $A_B$ and $A_V$ that
define $R_V$ depend on the SED of the star being observed.
Consequently, as shown in Fig.~\ref{fig:RV_implied} and by
\cite{McCall_only.2004}, one can place different sources behind the
same dust column and still obtain significantly different measurements
of $R_V$.
Moreover, as the broadband $A_V$ and $A_B$ do not depend linearly on
dust column density, the inferred $R_v$ depends also on the depth of
the dust column in front of the star.

Ideally we would like to have a dust extinction law that depends only
on the dust's intrinsic opacity $\kappa_\lambda$.
One possibility would be to use the value of $R_V$ that one would
measure if a vanishingly small amount of the dust were placed in front
of a standard star \citep{McCall_only.2004}, but this is unnecessarily
complicated.  Instead we follow \cite{MaizApellaniz_only.2013} and 
adopt the more straightforward quantity
\begin{align}
R_{5495} \equiv \frac{A_{5495}}{A_{4405}-A_{5495}},
\end{align}
where $A_{5495}$ and $A_{4405}$ are the monochromatic extinctions at
$5495$~\AA \, and $4405$~\AA \, respectively\footnote{$A_{5495}$ is
  sometimes denoted as $A_0$ \citep[e.g][]{Bailer-Jones_only.2011,
    Sale_Drew.2014}, following its use in
  \cite{Cardelli_Clayton.1989}. }.
This is designed to be similar to $R_V$, but, as $R_{5495}$ is defined
using monochromatic extinctions, it does not depend on the SED of the
star observed and will vary along a line of sight only if the grain
size distribution and therefore $\kappa$ varies.

With this choice of monochromatic wavelengths, the values of
$R_{5495}$ for the \cite{Fitzpatrick_only.2004} selection of
extinction laws are similar to the~$R_V$s they quote.
For example, their `$R_V=2.1,3.1,4.1$' curves give $R_{5495}=2.097,
3.056, 4.034$ respectively.
In Fig.~\ref{fig:RV_implied} we plot the $R_V$ implied by the 
$R_{5495}=3.056$ (`$R_{V}=3.1$') extinction law of 
\cite{Fitzpatrick_only.2004} for a range of SEDs along the main 
sequence and for various quantities of extinction.
As in \cite{McCall_only.2004}, it is apparent that there are
significant variations in $R_V$ along the main sequence, in addition
to smaller variations in response to increasing extinction.
We note that this procedure typically gives $R_V=3.1$ for late-B type
stars, a not unexpected result given that \cite{Fitzpatrick_only.2004}
used a sample of O,B and A stars to determine their extinction laws

\subsection{Selecting a wavelength for monochromatic extinctions}\label{sec:A_lambda}

Now that we have defined our extinction law, we can easily transform
monochromatic extinction given at one wavelength to any other
wavelength.
Therefore, we are free to choose the reference wavelength at which
monochromatic extinctions are defined.
When monochromatic extinctions have been used in earlier work, the
choice of wavelength has generally been made for reasons of
convenience.
For example, \cite{Hanson_Bailer-Jones.2014} and \cite{Sale_only.2012}
followed \cite{Bailer-Jones_only.2011} in adopting ``$A_0$'', the
monochromatic extinction at $5495$~\AA, chosen to enable easy use of
the \cite{Cardelli_Clayton.1989} extinction laws, which are anchored
at this wavelength.
In contrast, \cite{Sale_Drew.2009} used $A_{6250}$, the monochromatic
extinction at $6250$~\AA.
As this wavelength lies near the centre of the IPHAS $r$ band used in 
that study, the resulting measurements were less affected by 
variations in $R_{5495}$.

From~\eqref{eqn:Alambda_defn} we have that
\begin{align}
  \frac{dA_{\lambda}}{ds}(s) &= 1.086 \kappa_\lambda(s) \rho(s).
 \label{eqn:Alambda_diff}
\end{align}
If $\kappa$ did not change along a sightline, it would be trivial to
obtain the dust column density from $A_{\lambda}$ if $R_{5495}$, and
therefore $\kappa_\lambda$, were known.
In reality, however, we expect that the grain size distribution, and
consequently $R_{5495}$ and $\kappa$, will vary along lines of sight
as well as between them.
Instead we can look for a wavelength where $\kappa$ is approximately
independent of $R_{5495}$.
Examination of the \cite{Draine_only.2003} models indicates that
$\kappa_\lambda$ varies only weakly with changing dust grain
distribution at around $\lambda=4000$~\AA.
Therefore we have that
\begin{align}
A_{4000}(s)  \simeq  1.086 \kappa_{4000}\int_0^s \rho(s')\, ds' ,
\end{align}
where $\kappa_{4000}$ is the opacity at $4000$~\AA \, that
\cite{Draine_only.2003} quotes as $3.8 \times 10^{-3}{\rm m}^2 {\rm
  kg}^{-1}$ for his $R_V=3.1$ grain distribution.
So, adopting this $\lambda=$4000~\AA \, anchor point, we now have a
measure of extinction that -- to a reasonable approximation -- depends
on the column density of dust and is independent of variations in
opacity.
In order to facilitate comparisons to existing results, we note that,
if $R_{5495}=3.056$, then
\begin{align}
&  A_V \simeq 0.6929 A_{4000} + 0.0018 A_{4000}^2, \\
&  R_V \simeq 3.1 
\end{align}
for an A0V star.

Before proceeding further we pause to comment that, in the earlier
models of \cite{Weingartner_Draine.2001}, the opacity per unit dust
mass is a stronger function of $R_{5495}$ at $4000$~\AA, but is nearly
independent of $R_{5495}$ at approximately $8000$~\AA.
Therefore we advise that it may prove necessary to renormalise to
monochromatic extinction to a different wavelength in the future,
should improved dust models that contradict those of
\cite{Draine_only.2003} become available.
Changing this reference wavelength would have no effect on the methods
we propose below; the key point is that we choose a wavelength that 
minimises opacity variations, the actual wavelength itself is not 
directly important.

\subsection{Variation of broad-band extinction with dust column and
  stellar type}\label{sec:A_response}

\begin{figure*}
\includegraphics{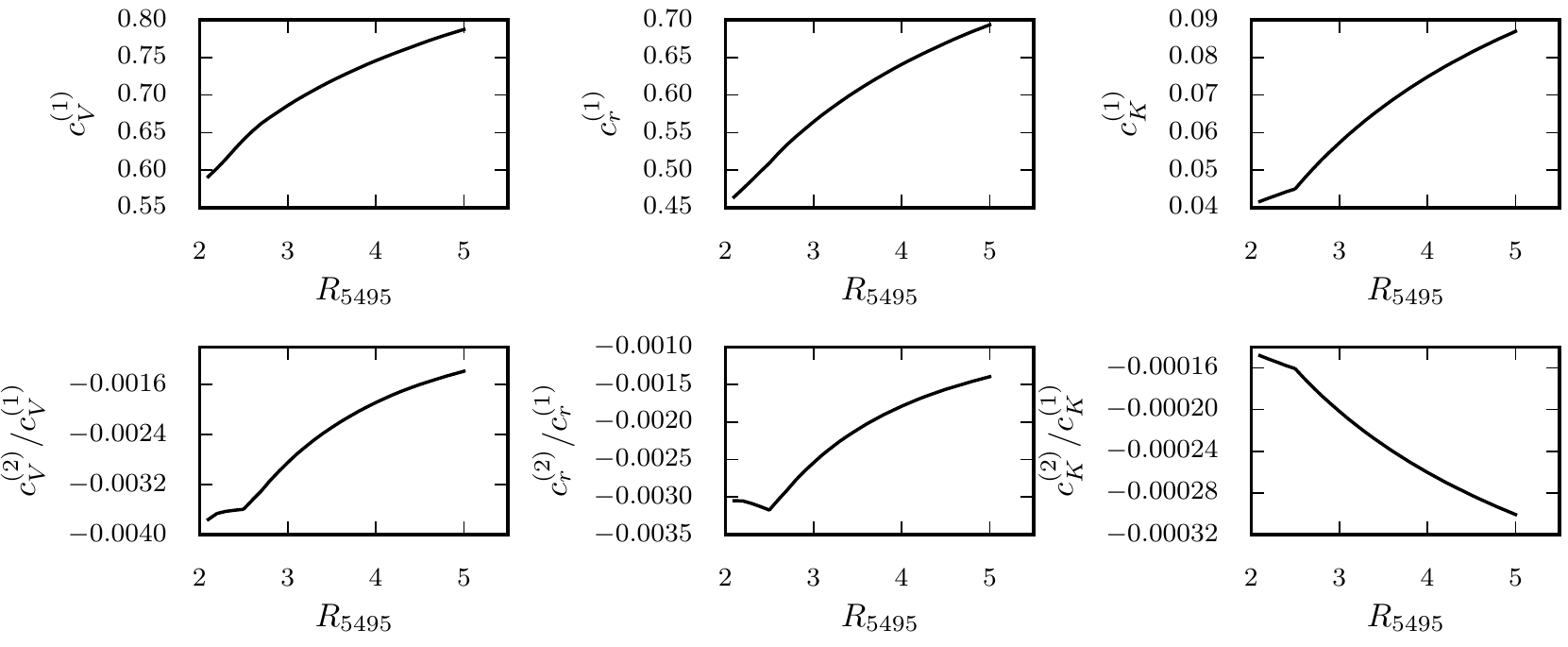}
\caption{Examples of the dependence of $\extnlin_X$ and $\extnquad_X$
  (equation~\ref{eqn:AXalphabeta}) on
  $R_{5495}$ for an approximate A0V star with $T_{\rm eff}=9600$~K and
  $\log g = 4.07$ and solar metallicity. The values in the left column
  are for the Johnson-Cousins $V$ band, as defined by
  \protect\cite{Bessell_only.1990}.
  The centre column is for the IPHAS $r$-band and the right column for
  the UKIDSS $K$-band as defined by \protect\cite{Hewett_Warren.2006}.
  \label{fig:ab_R5495}}
\end{figure*}

\begin{figure*}
\includegraphics{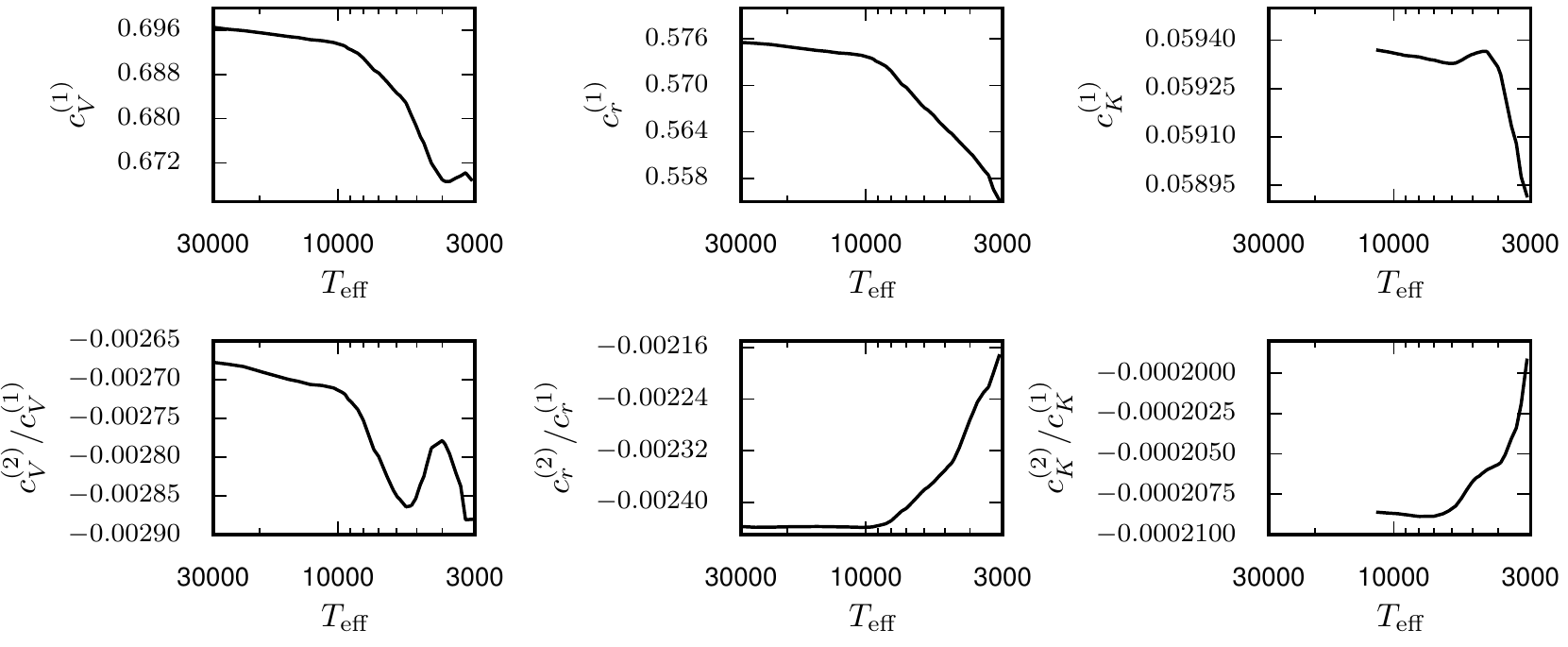}
\caption{Examples of the dependence of $\extnlin_X$ and $\extnquad_X$ on
  $T_{\rm eff}$ along the main sequence, as defined in $(T_{\rm eff},
  \log g)$ by \protect\cite{Straizys_Kuriliene.1981} for stars having
  solar metallicity.
  The columns are the same as Fig.~\ref{fig:ab_R5495}.
  \label{fig:ab_teff}}
\end{figure*}

\begin{figure*}
\includegraphics{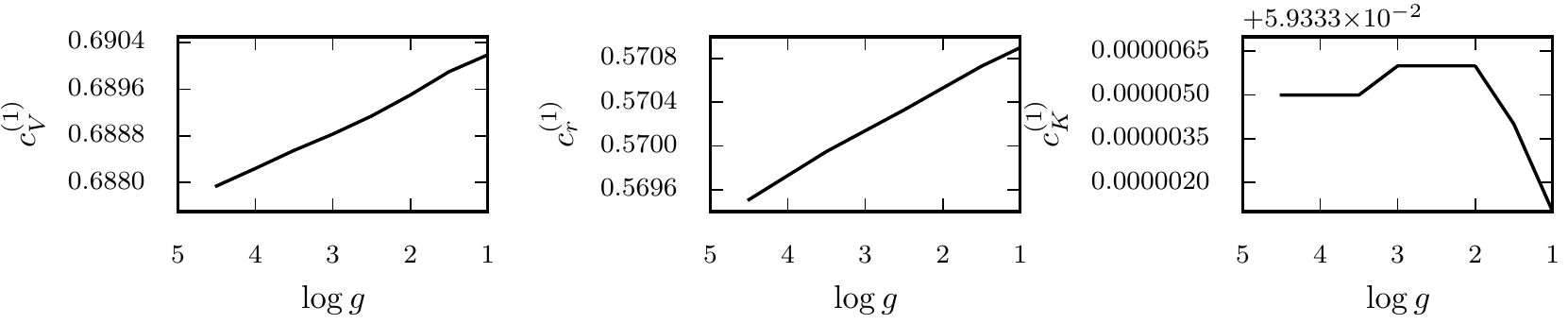}
\caption{Examples of the dependence of $\extnlin_X$ and $\extnquad_X$ on
  $\log g$.
  We fix $T_{\rm eff}=7000$~K and metallicity to solar.
  The columns are the same as Fig.~\ref{fig:ab_R5495}.
  \label{fig:ab_logg}}
\end{figure*}

\begin{figure*}
\includegraphics{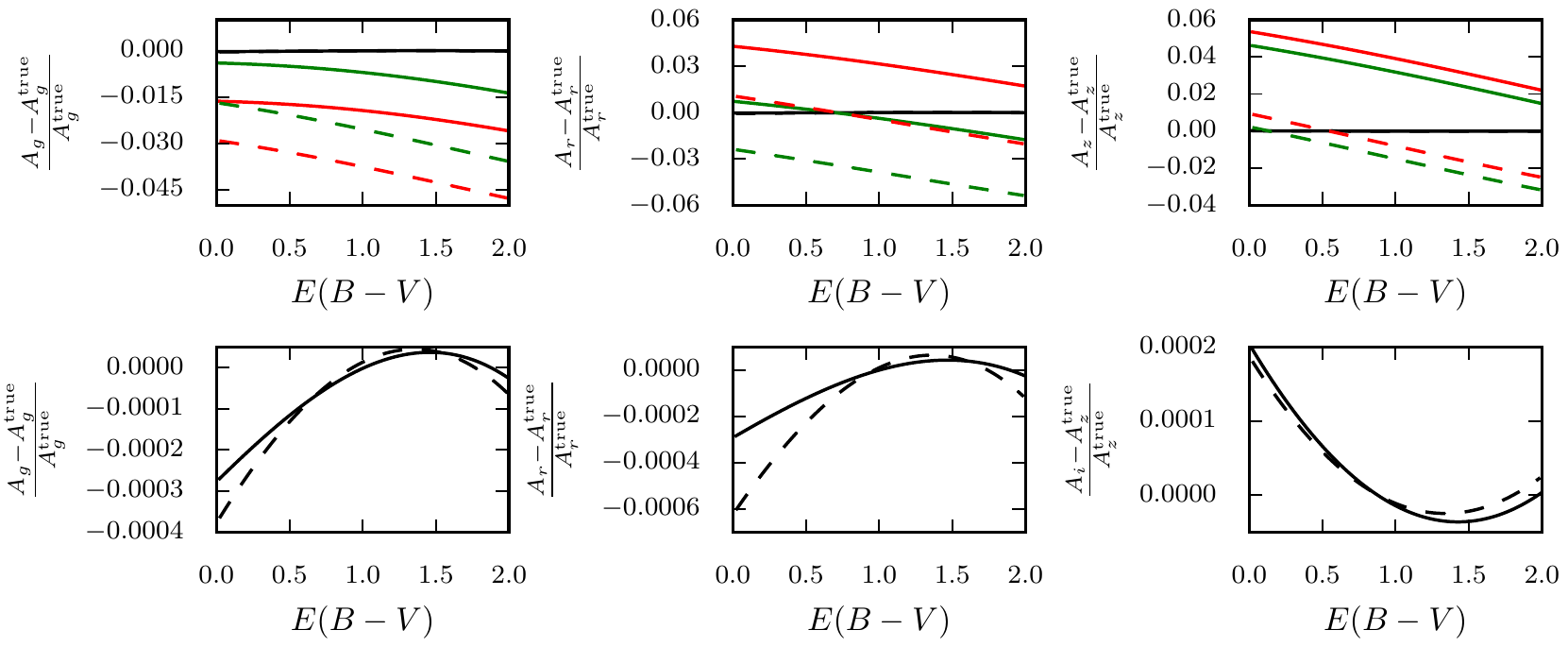}
\caption{  Comparison of responses to extinction in different SDSS 
bands using stars of type A0V (solid curves) and M3V (dashed) as 
examples.
Upper row:
  fractional errors in  $A_g$ (left), $A_r$ (middle) and $A_z$ (right)
  obtained assuming the $A_X/E(B-V)$ ratios given by
  \protect\cite{Schlegel_Finkbeiner.1998} (red),
  using $A_X/E(B-V)$ ratios from \protect\cite{Schlafly_Finkbeiner.2011} (green)
  and by using our fit~\eqref{eqn:AXalphabeta} (black).
  The solid (dashed) curves plot results for A0V (M3V) stars.
  Lower row: zoomed-in views of the fractional errors produced by our
  fits~\eqref{eqn:AXalphabeta}.
\label{fig:SFD_diff}}
\end{figure*}

We now examine how passband extinctions~$A_X$ depend on the
extinction law (i.e., $R_{5495}$) and stellar type.
We use a method similar to that employed in \cite{Sale_Drew.2009}.
For stars cooler than 12000~K we adopt the closest match from the Phoenix
library of synthetic spectra \citep{Husser_Wende-vonBerg.2013} to the 
star's SED.
The Phoenix library does not cover hotter stars, so, for 
$T_{\rm eff}>12000$~K we instead draw from the 
\cite{Munari_Sordo.2005} library.
As the \cite{Munari_Sordo.2005} library does not cover wavelengths 
redder than 10500~\AA, we do not provide results for the combination 
of hot stars and passbands with measured transmission beyond this 
wavelength.
Then we use equation~\eqref{eqn:AX_defn} to calculate~$A_X$, with
$\kappa_\lambda$ given by the appropriate \cite{Fitzpatrick_only.2004}
extinction law for the assumed~$R_{5495}$.

Clearly one does not want to repeat this procedure every time one
seeks to determine the model colours and apparent magnitude of some
star.  \cite{Casagrande_VandenBerg.2014} have suggested
precomputing the implied passband extinctions for all combinations of
model stars and extinction laws for a range of $A_{4000}$.  However,
storing all these extinctions for a reasonably dense isochrone library
and for a decent range of $R_{5495}$ and $A_{4000}$ would be
impractical.  Our approach is instead to use quadratic relations
\begin{align}
A_X=\extnlin_X A_{4000}+\extnquad_X A_{4000}^2
\label{eqn:AXalphabeta}
\end{align}
to fit the dependence of $A_X$ on $A_{4000}$ in the range $0\leq
A_{4000}<10$ for a grid of stellar parameters $(T_{\rm
  eff}, \log g, \feh)$ and extinction laws~$R_{5495}$.

As one might expect, the correction coefficients $\extnlin_X$ and
$\extnquad_X$ depend sensitively on the parameter $R_{5495}$ that sets
the shape of the extinction law (see, e.g., Fig.~\ref{fig:ab_R5495}).
Varying the effective temperature of the star is also significant, a
fact that is often overlooked (e.g., Fig.~\ref{fig:ab_teff}).  The
extent and form of the temperature dependence varies from band to
band, and is governed by the position of the band relative to the peak
of the star's spectrum.

Varying $\log g$, $\feh$ or $\afe$ affects $A_X$
much less than variations in $T_{\rm eff}$.
For example, Fig.~\ref{fig:ab_logg} plots the dependence of
$\extnlin_X$ and $\extnquad_X$ on $\log g$, showing that variations in the
latter produce changes that are more than an order of magnitude
smaller than those caused by varying~$T_{\rm eff}$.
The response to changes in $\feh$ or $\afe$ is
significantly smaller still; only in $u$ or similar bands, where
variations in $\feh$ most strongly affect spectra, can the chemical
composition of the star measurably affect the effect of reddening and
then typically only at a level comparable to $\log g$.

Therefore, in general we approximate $\extnlin_X$ and $\extnquad_X$ as
functions of $R_{5495}$ and $T_{\rm eff}$ only, neglecting the
dependence on $\log g$, $\feh$ and $\afe$.
In appendix~\ref{app:abs} we provide tables of $p$ and $q$
for a variety of different popular photometric systems for a range of
$T_{\rm eff}$ and~$R_{5495}$.
We encourage readers to adopt this calibration when considering
extinction for stars within our Galaxy, since it accounts for SED and
extinction law variation as well as the nonlinear response of
extinction in any given photometric band due to increasing dust column.

In the absence of a proper calibration, such as that in
appendix~\ref{app:abs}, many have turned to table~6 of
\cite{Schlegel_Finkbeiner.1998} to convert between extinctions in
different bands.
We warn that this table of relative extinctions was not intended to be
used for stars within the Galaxy, but rather to deredden photometry 
of galaxies behind the relatively sparse dust columns that
characterise high Galactic latitudes.
As such their table was calibrated using an elliptical galaxy as a
source in the limit of small extinctions.
Therefore, the calculated ratios are not suitable for detailed use
when considering stars subject to significant extinction.

In Fig.~\ref{fig:SFD_diff} we show the fractional errors that arise due
to assuming parametrizations of extinction given here and those that 
result from assuming the $A_X / E(B-V)$ ratios given by 
\cite{Schlegel_Finkbeiner.1998} and \cite{Schlafly_Finkbeiner.2011}.
The errors due to the calibration we propose here are extremely small, 
generally less than 0.1\%. 
In contrast, the error that arises by following the 
\cite{Schlegel_Finkbeiner.1998} or \cite{Schlafly_Finkbeiner.2011} 
ratios is frequently on the order of 5\%.

\section{Marginal likelihood of distance and
  extinction}\label{sec:prob}

Having discussed how best to model the effects of extinction, we now
turn to the problem of estimating distances and extinctions to
individual stars in situations where we are not interested in the
details of each star's spectral type.
This problem occurs when constructing three-dimensional extinction
maps from stellar catalogues \citep[e.g.,][]{Sale_Magorrian.2014}.

It is convenient to replace the distance~$s$ by the distance modulus
\begin{equation}
\mu \equiv 5 \log_{10} (s/10\hbox{pc})
\end{equation}
and the extinction $A_{4000}$ by its logarithm,
\begin{equation}
a_{4000} \equiv \ln A_{4000}.
\end{equation}
As $s>0$ and we assume that $A_{4000} > 0$, it is sensible to consider
the logarithms of both values, since both $\mu$ and $a_{4000}$ span
the entire real line.
In addition, the use of the distance modulus is sensible, as
uncertainties on it are often approximately Gaussian.
Meanwhile, the use of $a_{4000}$ is largely motivated by the fact that
in \cite{Sale_Magorrian.2014} we place a Gaussian random field prior
on it to create an extinction map.

The particular problem we address is the following.
Given a set of observations~$\vy$ of some star, we would like to
compute the marginal likelihood
\begin{equation}
  \label{eqn:margxlik}
\pr(\vy|\mu,a_{4000},R_{5495},\transfermodel,\galaxymodel)=
\int\pr(\vy|\mu,a_{4000},R_{5495},\vx,\transfermodel)
\pr(\vx|\mu,\galaxymodel)\,\d\vx
\end{equation}
of the distance modulus~$\mu$ and log-extinction~$a_{4000}$ to the star by
marginalising the star's intrinsic parameters~$\vx$, which include 
its mass, age, metallicity and so on.
We assume a mix of stellar populations~$\galaxymodel$, which specifies the
(possibly position-dependent) prior $\pr(\vx|\mu,\galaxymodel)$
on~$\vx$.\footnote{Recall that $\pr(\vx|\mu,\galaxymodel)$ is actually
  $\pr(\vx|\mu,l,b,\galaxymodel)$, as we have suppressed the dependence on
  the line of sight $(l,b)$.}
When the observations~$\vy=\vy_{\rm phot}$ are limited to the star's
photometric apparent magnitudes the likelihood $\pr(\vy_{\rm
  phot}|\mu,a_{4000},R_{5495},\vx,\transfermodel)$ is straightforward to
calculate using the extinction model~$\transfermodel$ described in
Section~\ref{sec:parametrizing}.
If one has independent additional data $\vy_{\rm other}$, such as a
spectroscopic metallicity or a trigonometric parallax, then the
likelihood becomes
\begin{equation}
  \begin{split}
&\pr(\vy|\mu,a_{4000},R_{5495},\transfermodel,\galaxymodel)\\
&=
\pr(\vy_{\rm phot}|\mu,a_{4000},R_{5495},\transfermodel,\galaxymodel)
\pr(\vy_{\rm other}|\mu,\galaxymodel),
  \end{split}
\end{equation}
in which $\pr(\vy_{\rm other}|\mu,\galaxymodel)$ is (usually) independent of
extinction and is relatively easy to treat.
In the following we ignore $\pr(\vy_{\rm other}|\mu,\galaxymodel)$ and assume
that $\vy=\vy_{\rm phot}$ only.

The dependence of the marginalised likelihood $\pr(\vy_{\rm
  phot}|\mu,a_{4000},R_{5495},\transfermodel,\galaxymodel)$ on $(\mu,a_{4000}, R_{5495})$ is
usually difficult to predict directly from the observed~$\vy_{\rm
  phot}$.
We might reasonably expect that it will typically have two maxima -- one 
that corresponds to the star being on the main sequence, the other to 
the giant branch --
but the locations $(\mu,a_{4000}, R_{5495})$ and extent of these maxima cannot be
found without some exploration of the $(\mu,a_{4000}, R_{5495})$ space.
We use an MCMC algorithm to carry out this exploration, and then fit a
simple mixture model to the $(\mu,a_{4000}, R_{5495})$ dependence of the
likelihood function.

The marginal likelihood depends on a calibration $\transfermodel$, 
that includes the extinction calibration discussed in 
section~\ref{sec:parametrizing} in addition to a set of isochrones 
rendered in the appropriate filter system.
In this paper we employ Padova isochrones \citep{Bressan_Marigo.2012}, 
which use bolometric corrections calculated from \textit{ATLAS9}
\citep{Castelli_Kurucz.2003} model spectra.
To investigate the extent of the systematic error stemming from the 
use of \textit{ATLAS9} derived bolometric corrections, we have repeated the 
tests of section~\ref{sec:fit} but with bolometric corrections derived 
from Phoenix model spectra.
We found that the posterior expectations of distance modulus and 
extinction typically change by $\lesssim 0.02$.
We caution that, although this uncertainty may appear small, as it is 
systematic it will potentially have a measurable impact on extinction 
maps.

\subsection{Sampling the likelihood function}

Although the marginalised likelihood~\eqref{eqn:margxlik} is a
function of $(\mu,a_{4000},R_{5495})$, it is not a probability density in these
parameters and therefore cannot directly be sampled using MCMC methods.
So, we begin by using Bayes' theorem to express it as
\begin{equation}
  \begin{split}
&\pr(\vy|\mu,a_{4000},R_{5495},\transfermodel,\galaxymodel)\\
&=
\pr(\vy|\transfermodel,\galaxymodel)
\frac{\int\pr(\mu,a_{4000},R_{5495},\vx|\vy,\transfermodel,\galaxymodel)\,\d\vx}
{\pr(\mu,a_{4000},R_{5495}|\galaxymodel)},
  \end{split}
\label{eqn:lik2post}
\end{equation}
in which the prior on
$(\mu,a_{4000},R_{5495})$ that appears in the denominator is given by
\begin{equation}
\label{eqn:lik2postprior}
\pr(\mu,a_{4000},R_{5495}|\galaxymodel)
=\int   \pr(\mu,a_{4000},R_{5495},\vx|\galaxymodel)\,\d\vx,
\end{equation}
which is completely determined by our choice of galaxy
model~$\galaxymodel$.
Now the posterior
\begin{equation}
  \begin{split}
&\pr(\mu,a_{4000},R_{5495},\vx|\vy,\transfermodel,\galaxymodel)\\
&=
\frac{\pr(\vy|\mu,a_{4000},R_{5495},\vx,\transfermodel,\galaxymodel)
\pr(\mu,a_{4000},R_{5495},\vx|\galaxymodel)}
{\pr(\vy|\transfermodel,\galaxymodel)}
  \end{split}
\label{eqn:MCMCposterior}
\end{equation}
that appears within the integral in the numerator
of~\eqref{eqn:lik2post} can be sampled using any convenient MCMC
method.
The likelihood
$\pr(\vy|\mu,a_{4000},R_{5495},\vx,\transfermodel,\galaxymodel)$ is
easy to calculate and the normalising factor
$\pr(\vy|\transfermodel,\galaxymodel)$ is important only if we want to compare
models with different population mixes~$\galaxymodel$ or dust
properties $\transfermodel$.

\subsubsection{The prior $\pr(s,A_{4000},R_{5495},\vx|\galaxymodel)$}

In the examples that follow we adopt a prior $\pr(\vx|\mu,
\galaxymodel)$ on the stellar parameters that comes from an
intentionally simple Galactic model~$\galaxymodel$.
We include in $\galaxymodel$ a Salpeter IMF and a constant star
formation history.
We model metallicity variations as being Gaussian with a standard
deviation of 0.2~dex and a mean that declines by 0.06~dex~kpc$^{-1}$
with Galactocentric radius, following \cite{Luck_Lambert.2011},
normalised to solar metallicity at the solar circle.
We could also impose a vertical metallicity gradient, but opt not to
do so here since the stars we use as examples later in this section
lie very close to the Galactic mid-plane.

In the light of~\eqref{eqn:lik2post}, we are free to use any
convenient prior on the parameters $\mu$, $a_{4000}$ and $R_{5495}$
provided only that it is sufficiently broad to cover plausible regions
of parameter space.
We adopt a flat prior in all three parameters, but with $R_{5495}$
limited to the range $2.1\lesssim R_{5495} \lesssim 5.5$ over which
the \cite{Fitzpatrick_only.2004} reddening laws are available.
When written out, the prior in~\eqref{eqn:MCMCposterior} is
\begin{equation}
\begin{split}
  \pr(\mu,a_{4000},R_{5495},\vx|\galaxymodel) &\propto
\pr(\vx|\mu, \galaxymodel)\pr(\mu,a_{4000}, R_{5495} ), \\
\pr(\vx|\mu, \galaxymodel) &\propto \mathcal{M}^{-2.35} \mathcal{N}(\mh | 0.06(\mathcal{R}_{\odot}-\mathcal{R}) , 0.2), \\
\pr(\mu,a_{4000}, R_{5495} ) & \propto \begin{cases} 1, &\mbox{if } 2.097 \leq R_{5495} < 5.402, \\
0, &\mbox{otherwise}, \end{cases}
\end{split}
  \label{eqn:MCMCposterior2}
\end{equation}
where $\mathcal{M}$ is the initial mass of the star, $\mh$ its
metallicity, $\mathcal{R}$ the Galactocentric radius of the star
implied by the distance modulus~$\mu$ and Galactic coordinates
$(l,b)$, $\mathcal{R}_{\odot}$ the Galactocentric radius of the sun, and
$\mathcal{N}(\cdot|\cdot,\cdot)$ denotes a normal distribution.

\subsubsection{The choice of MCMC scheme}

\begin{figure*}
\includegraphics{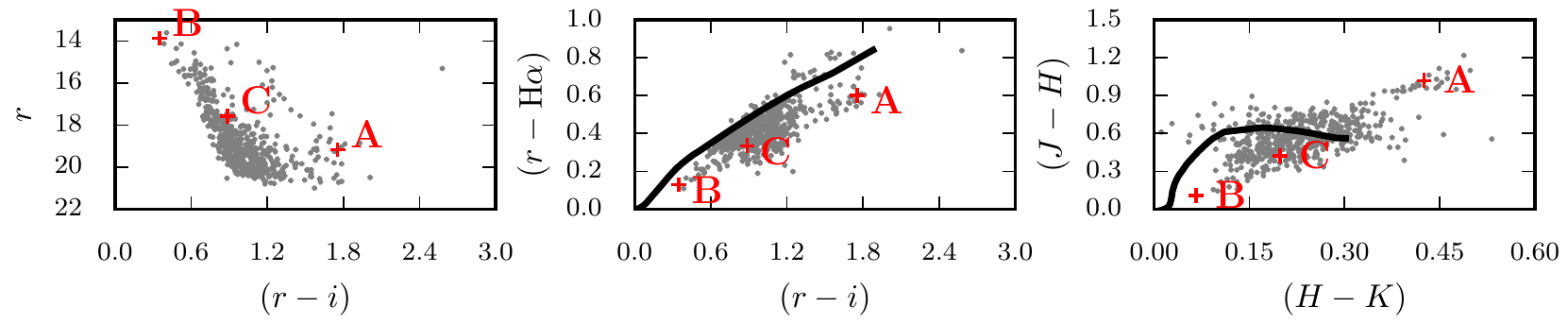}
\caption{Colour-colour diagrams in the IPHAS and UKIDSS-GPS systems of our sample catalogue. The three stars studied in detail are marked with red crosses, labelled with their corresponding letter, where: A is IPHAS2 J211210.70+482106.8, B is IPHAS2 J211225.40+481927.6  and C is IPHAS2 J211223.10+481656.4, Solid lines show unreddened main sequences. Only stars that appear in all six bands and are flagged as stellar in both surveys are shown. 
\label{fig:photom}}
\end{figure*}

\begin{figure*}
\includegraphics{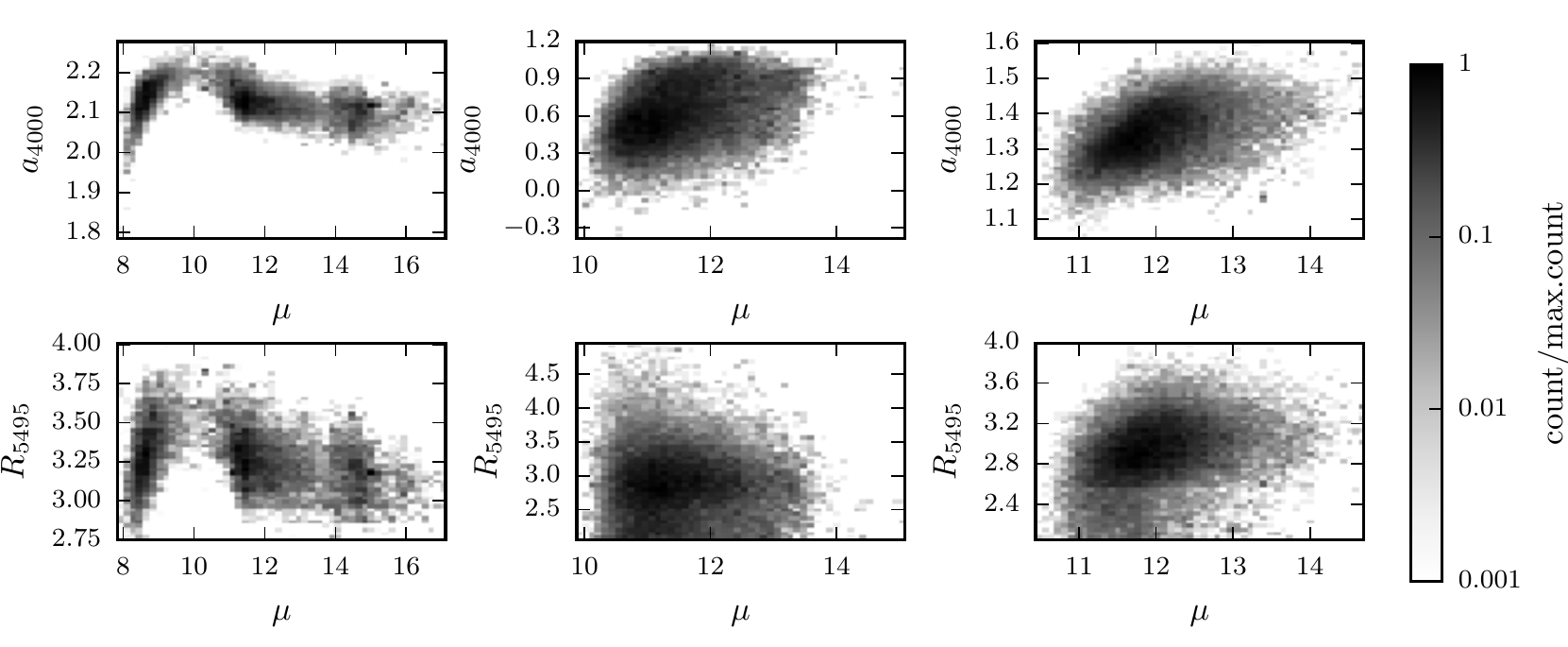}
\caption{Example marginal likelihoods shown for stars A (left), B (centre) and C (right). We have binned the MCMC samples only for the purposes of producing histograms.
\label{fig:marg_ex}}
\end{figure*}

There are many MCMC algorithms that could be used to sample the
pdf~\eqref{eqn:MCMCposterior}.
We use the affine invariant ensemble sampler of
\cite{Goodman_Weare.2010} as implemented in \textsc{emcee}
\citep{Foreman-Mackey_Hogg.2013}.
This algorithm employs a collection of `walkers' that explore the
parameter space.
At each iteration each walker attempts to move some distance along the
vector towards another randomly chosen walker.
We initialise the sampler with an array of 100 walkers positioned
along the main sequence and red giant branch in $(T_{\rm eff}, \log
g)$ space, with $\mh$ drawn from the prior distribution, $R_{5495}$ 
drawn from a uniform distribution on the full range of 
\cite{Fitzpatrick_only.2004} reddening laws and $\mu$ and $a_{4000}$ 
chosen to be the maximum likelihood values given all the other 
parameters.
However, when running this algorithm it became clear that groups
of walkers would occasionally become stuck in islands of low
probability, with the relatively high dimensionality making it
difficult for them to transition out to higher probability regions.
We therefore adopt a similar approach to \cite{Hou_Goodman.2012} and
`prune' the set of walkers at the end of burn-in, moving some walkers
when a disproportionally large number are stuck in islands of low
probability.

Our general schema then consists of using 100 walkers, with a burn-in
of 1000 iterations, of which the last 100, thinned by a factor of 10,
are used to facilitate the pruning.
After the pruning, we then iterate for a further 9000 iterations to
obtain our final MCMC chain, thinning the chain by a factor of 10.
The thinned chain typically has an autocorrelation length of around 1,
implying an autocorrelation length of roughly 10 for the unthinned chain,
with a total sample size of 90,000 and an effective sample size of
roughly 45,000.
Although the moments of the distribution could be found to sufficient
precision with a much smaller effective sample size, capturing some of
the detail in the posterior requires such a large sample.
We can then perform the integration in~\eqref{eqn:lik2post} by simply
ignoring all parameters other than $\mu$, $a_{4000}$ and $R_{5495}$ in 
the MCMC chain.  Dividing the result by our (trivial) prior gives the
desired marginal likelihood
$\pr(\vy|\mu,a_{4000},R_{5495},\transfermodel,\galaxymodel)$.

We demonstrate how we obtain the marginal likelihoods using photometry 
from IPHAS \citep[INT/WFC photometric $\Halpha$ survey of the 
northern Galactic plane;][]{Drew_Greimel.2005, Barentsen_Farnhill.2014}
and UKIDSS-GPS \citep[UKIRT infrared deep sky survey - Galactic plane 
survey;][]{Lucas_Hoare.2008}.
In particular we use a crossmatched catalogue that covers 
$5\arcmin \times 5\arcmin$ centred on $(l,b) = (90.04, -0.04)$.
We use a $1$~arcsec matching radius and only stars flagged as stellar 
in both surveys are included. 
We show In Fig.~\ref{fig:photom} colour-colour plots of this 
catalogue. 
From this catalogue we select three stars to concentrate on: 
IPHAS2 J211210.70+482106.8, IPHAS2 J211225.40+481927.6  and 
IPHAS2 J211223.10+481656.4, which we label as stars A,B and C 
respectively.
These span a range of colours and apparent magnitudes.
We add, in quadrature, to the stated photometric uncertainties an 
additional factor of 2\% to account for systematic uncertainties, such 
as those on the photometric zero points.
This additional factor dominates the uncertainty budget for stars B 
and C and makes an important contribution for star A.

Two-dimensional histograms of the marginal likelihoods obtained for
the three stars are displayed in Fig.~\ref{fig:marg_ex}.
It is apparent that, as in \cite{Green_Schlafly.2014}, some 
exhibit complicated shapes, largely due to to the irregular
shape of the stellar locus in colour--magnitude space.
In particular, star A could be either a main sequence star or on the
red giant branch: from its photometry alone we are unable to make a
distinction.
On the other hand, qualitative examination of the colour--magnitude
diagram in Fig.~\ref{fig:photom} indicates that the star should be on
the giant branch, due to its position in a redder sequence
\citep{Sale_Drew.2009}.
However, this qualitative analysis has been implicitly conditioned
upon the photometry of all the stars in the catalogue -- we would not
have been able to identify a red sequence if we only had the
photometry of star A.
In contrast, the likelihood in Fig.~\ref{fig:marg_ex} is conditioned
upon only the photometry of star A.
In order to condition it on the entire photometric catalogue we
require a method such as that of \cite{Sale_Magorrian.2014} (see in
particular their equation 19), in which case the construction of the
extinction map would break the degeneracy between the main sequence
and red giant branch.
Both stars B and C appear too hot to be on the red giant branch.

Although we have assumed a flat prior on $R_{5495}$, the combination
of optical and near-infrared photometry has enabled us to narrow the
range of possible extinction laws \citep[see
also][]{Berry_Ivezic.2012}.
If our data had not constrained $R_{5495}$ our uncertainties on both
$\mu$ and $a_{4000}$ would have been increased, since $R_{5495}$ is
covariant with both.
We note that the uncertainties on $R_{5495}$ depend, to a large 
degree, on the number and wavelength range of the photometric bands 
employed: if, as in \cite{Berry_Ivezic.2012}, we had used SDSS data in 
place of IPHAS we would have 8 bands instead of 6 and the bluer 
coverage of the $u$ and $g$ bands and so should be able to achieve 
more precise estimates of $R_{5495}$.

\subsection{Fitting a mixture model to the marginalised likelihood function}\label{sec:fit}

\begin{figure*}
\includegraphics{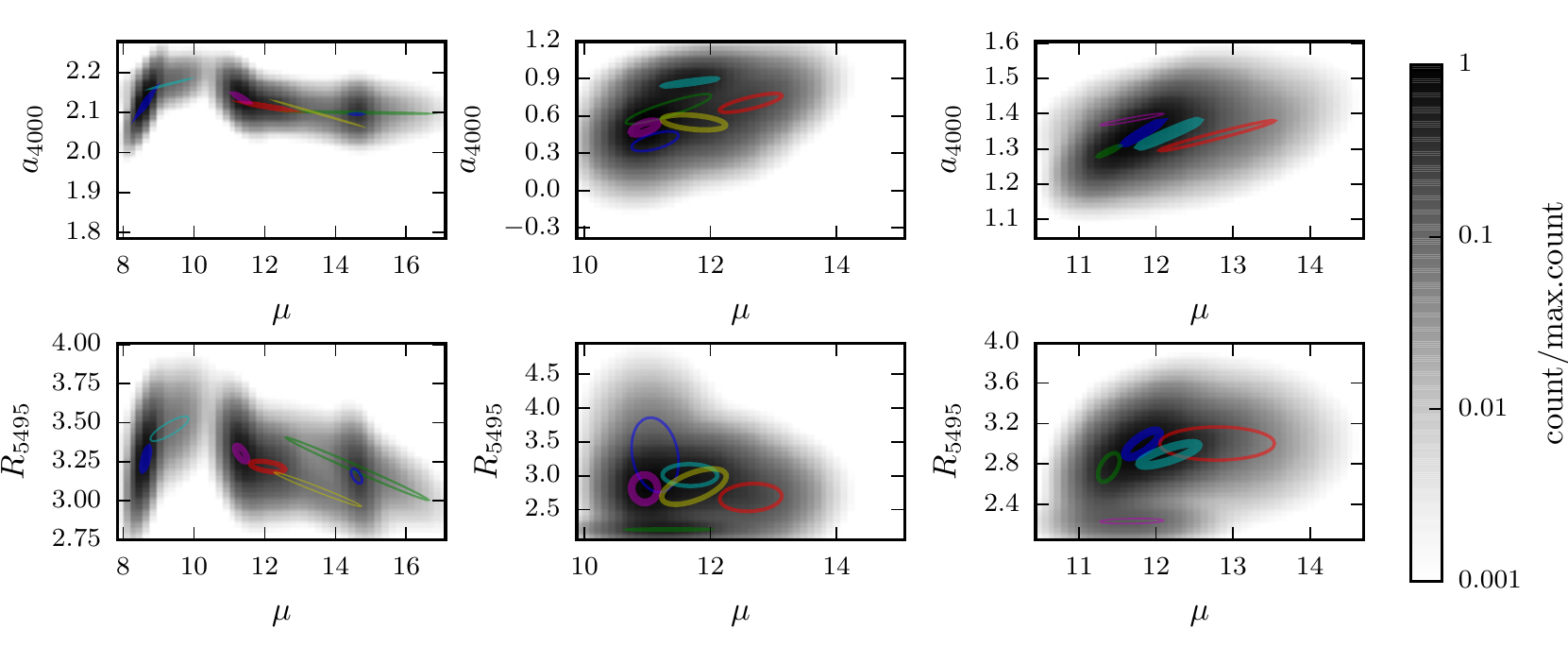}
\caption{Examples of the Gaussian mixture approximations, plotted in histograms to match Fig.~\ref{fig:marg_ex}. The coloured ellipses show the 2-$\sigma$ contours of each of the Gaussian mixture model components, with the width of the ellipses' curves linearly increasing with the weight of the corresponding component in the Gaussian mixture model. As in Fig.~\ref{fig:marg_ex} star A is in the left hand column, B in the centre and C on the left.
\label{fig:gauss_ex}}
\end{figure*}

\begin{figure*}
\includegraphics{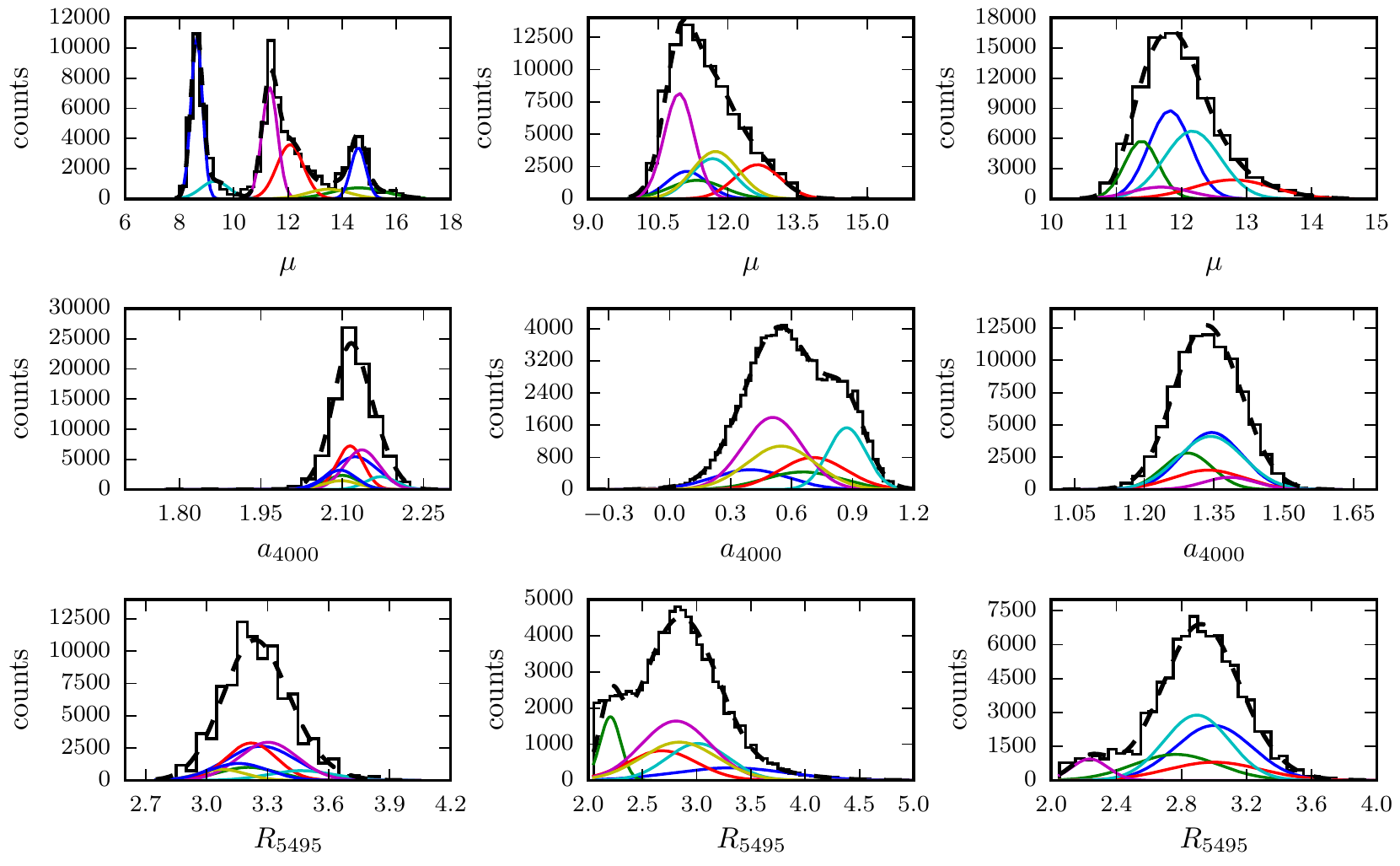}
\caption{Examples of 1D marginal likelihoods for each star shown fitted with Gaussian mixtures. The black solid line shows a histogram of the MCMC samples. As in Fig.~\ref{fig:marg_ex} we have only performed the binning of the MCMC samples to produce the plotted histograms. The coloured lines show the contribution of each of the components in the Gaussian mixture model, whilst the dashed black line shows the total one dimensional marginal likelihood implied by the Gaussian mixture model. Once again star A is in the left hand column, B in the centre and C on the left.
\label{fig:1d_mu}}
\end{figure*}

Having explored the $(\mu,a_{4000},R_{5495},\vx)$ posterior, we carry
out the marginalisation of~$\vx$ in the numerator
of~\eqref{eqn:lik2post} by simply ignoring the $\vx$ values returned
by the sampler and focusing only on the distribution of the
$(\mu,a_{4000},R_{5495})$ samples.  
These samples are drawn from the marginal likelihood
function~\eqref{eqn:margxlik} weighted by the
prior~\eqref{eqn:lik2postprior} of our assumed
galaxy model~$\galaxymodel$.

Although it would be possible to use the full set of samples from the
MCMC chain (re-weighted to account for the prior) as our description of
the marginalised likelihood, this is far from ideal.
The chains are long.
Therefore the cost of storing them is high, all the more so when one
considers that, when constructing coherent maps of extinction or
stellar density, one will typically want to use marginalised
likelihoods for many stars simultaneously.
The data volumes can be reduced by thinning the MCMC chain (i.e., by
removing all but every $n^{\rm th}$ entry).
Drastic thinning would enable the data volumes to be manageable, even
with a large catalogue of stars.
The cost associated with thinning, however, is that it reduces the
ability of the chain to represent the true underlying marginalised
likelihood function, particularly in the relatively low likelihood
regions.
This can be a key problem if the likelihood is then fed into a
hierarchical model, as in \cite{Sale_Magorrian.2014}, where data from
other stars suggests that the distance or extinction to this star may
lie in such a low probability region.
For example, if the range of possible extinctions to a particular star
were constrained by other nearby stars to a region that is only
sampled by a single point in the thinned MCMC chain, the resultant
marginal posterior distribution of extinction to this star will take
the form of a delta function and the uncertainty would therefore be
drastically underestimated.

At the other extreme, the most common solution to this problem is to
simply report the mean and covariance matrix of the likelihood
function.
But doing this does not pass on any detailed information about the
shape of the likelihood function, which, as demonstrated by
Fig.~\ref{fig:marg_ex}, may well be somewhat irregular.
In particular it will not reveal multimodality, as might be the case 
if there are two peaks in the likelihood corresponding to the 
observed star being on the main sequence or on the giant branch.

An alternative is to describe the likelihood function using some
mixture of simple distributions.
For example, \cite{CarrascoKind_Brunner.2014} depict the posterior
distributions of photometric redshifts to galaxies using a mixture of
Gaussians and Voight profiles.
We instead fit a mixture of trivariate Gaussians to the marginalised
posterior $\pr(\mu,a_{4000},R_{5495}|\vy,\transfermodel,
\galaxymodel)$.
As we assume a flat prior on $(\mu,a_{4000},R_{5495})$, this is
equivalent to fitting Gaussians to the marginalised likelihood
function $\pr(\vy|\mu,a_{4000},R_{5495},\transfermodel,
\galaxymodel)$.
So, writing $\vtheta\equiv(\mu,a_{4000},R_{5495})$, our goal is to fit a
function
\begin{equation}
\pr(\vtheta|\vy,\galaxymodel,\transfermodel)\approx  \sum_{k=1}^Kw_k\mathcal{N}(\vtheta|\vm_k,\Cmat_k),
\label{eqn:gaussianmixture}
\end{equation}
to our MCMC sample $(\vtheta_1,...,\vtheta_N)$ by adjusting the weights $w_k$,
means $\vm_k$ and covariances $\Cmat_k$ of the Gaussians on the
right-hand side, along with their number~$K$.

Before explaining our procedure for carrying out the fitting, we note
that using Gaussians here has the key advantage that one can often
carry out further marginalisation analytically.
An example of this is given in Section~4.2 of
\cite{Sale_Magorrian.2014}, in which the distances and extinctions to
individual stars were marginalised in order to obtain the pdf of the
parameters describing the large-scale extinction distribution.
Similarly, having fit the trivariate Gaussian mixture model above, one
could later decide to take the prior $\pr(R_{5495}|\galaxymodel)$ to
be Gaussian with a mean and standard deviation from e.g.
\cite{Fitzpatrick_Massa.2007} and then marginalise $R_{5495}$
analytically to obtain the marginal likelihood
$\pr(\vy|\mu,a_{4000},\transfermodel, \galaxymodel)$.
This new, two-dimensional marginal likelihood would still be expressed
as a sum of Gaussians.

\subsubsection{Fitting a Gaussian mixture model with $K$ components}

One way of addressing the problem of fitting the Gaussian mixture
model~\eqref{eqn:gaussianmixture} to the MCMC sample would be by
modelling the latter as a Dirichlet process mixture of Gaussians.
Our goal here though is not to consider all possible Gaussian-mixture
descriptions of the MCMC chain, but instead to obtain a {\it single},
compact, ``best'' description of the marginalised likelihood.
Generally a single Gaussian will not describe the marginalised
posterior well, but a mixture of two or more Gaussians will do better.

A simple and robust approach is to iterate of a range of possible~$K$.
For each $K$ we use the expectation--maximization (EM) algorithm, as
implemented in \texttt{scikit-learn} \citep{Pedregosa_Varoquaux.2011},
to find the parameters $(w_k,\vm_k,\Cmat_k)$ of each of the $K$
Gaussians that maximise the likelihood
\begin{equation}
\mathcal L_K\equiv  \prod_{n=1}^N\sum_{k=1}^Kw_k\mathcal N(\vtheta_n|\vm_k,\Cmat_k),
\end{equation}
subject to the constraint that $\sum_kw_k=1$.
The EM algorithm functions by introducing $N\times K$ new latent variables
$\{z_{nk}\}$ that allow the awkward product of sums in this likelihood
to be rewritten as the easier-to-handle sum of products
\begin{equation}
\mathcal L_K\equiv
\sum_{\{z_{nk}\}}\prod_{n=1}^N\prod_{k=1}^K
\left[w_k\mathcal N(\vtheta_n|\vm_k,\Cmat_k)\right]^{z_{nk}}.
\end{equation}
The new variables $z_{nk}$ indicate the probability that MCMC sample
$n$ was drawn from Gaussian~$k$.
The algorithm proceeds by alternately updating the latent membership
probabilities $\{z_{nk}\}$ holding $\{w_k,\vm_k,\Cmat_k\}$ fixed, then,
for this choice of $\{z_{nk}\}$, finding the $\{w_k,\vm_k,\Cmat_k\}$
that maximise the likelihood.
We initialise the EM run with the means of the components given by the
mean of the MCMC sample and with diagonal covariance matrices with the
variance for each parameter being the corresponding variance from the
MCMC sample.
The EM algorithm is then run for 100 iterations to find optimal values
of $\{w_k,\vm_k,\Cmat_k\}$.

\subsubsection{How many components $K$?}

Having obtained maximum likelihoods for $K=1,2,3,...$ the question
then becomes one of deciding how many Gaussians are actually
justified.
For example, if we chose $K\ge N$ (i.e., there are as many Gaussians
as there are MCMC samples), then the likelihood would be unbounded:
simply centre one Gaussian on each point from the MCMC sample and let
its covariance shrink to zero.
We would like to avoid fitting the shot noise in our MCMC samples like
this, or, more practically, requiring such a large number of Gaussians
that they cause data volume problems.

A natural way of comparing models with different components would be
to adopt uninformative priors on $\{w_k,\vm_k,\Cmat_k\}$ and to
marginalise the likelihoods $\mathcal L_k$ to obtain the marginal
likelihoods $\pr(\{\vtheta\}|K)$ for each~$K$.
These $\pr(\{\vtheta\}|K)$ could be estimated by a variational Bayes
method \cite[see, e.g., Appendix~C of][]{Magorrian_only.2014}, but
doing so would be overkill for our present purposes.
As a straightforward alternative, we instead employ the Bayesian
Information Criterion \citep[BIC,][]{Schwarz_only.1978}
\begin{align}
{\rm BIC} =  -2 \ln\hat{\mathcal L}_K + (10 K -1) \ln N,
\end{align}
where $\hat{\mathcal L}_K$ is the maximum likelihood of the
$K$-component Gaussian mixture model, as found by the EM algorithm.
The second term in this expression acts as a penalty on the number of
components, with the $(10K-1)$ factor accounting for the number of
free parameters in a $K$-component trivariate Gaussian mixture model:
$3K$ numbers are needed to specify the means~$\vm_k$, $6K$ for the
symmetric covariance matrices $\Cmat_k$, and $K-1$ for the
weights~$w_k$.
Our favoured model is simply the one that minimises the value of ${\rm
  BIC}$.
We find that this minimum is typically achieved for mixtures having $K
\sim 5$ Gaussians.

Fig.~\ref{fig:gauss_ex} shows our Gaussian-mixture approximations to
the MCMC-sampled marginal likelihoods of Fig.~\ref{fig:marg_ex}.
As it is difficult to compare these 2D projections by eye, in
Fig.~\ref{fig:1d_mu} we also show 1D projections of both the MCMC
chain and our Gaussian-mixture fits.
The Gaussian mixture model provides a good, compact descriptions of
the the MCMC samples.

The one area in which we find that the mixture model fails to perform
well is when $R_{5495}$ takes on values close to the cut-offs imposed
by the range covered by the \cite{Fitzpatrick_only.2004} extinction
curves; our marginalised likelihoods fall sharply to zero at these
extreme values, a behaviour which the Gaussian mixture has difficulty
reproducing.
However, we note that the impact of such issues will be dramatically
reduced by the imposition of any sensible prior on $R_{5495}$.
For example, one could place a simple Gaussian prior on $R_{5495}$
with a mean and variance taken from e.g.
\cite{Fitzpatrick_Massa.2007}.
Under such a prior the probability of the problematic extreme values
of $R_{5495}$ would be very low and so the issues related to the fit
would become essentially irrelevant.

\subsection{The quality of the Gaussian mixture model approximation}

\begin{figure*}
\includegraphics{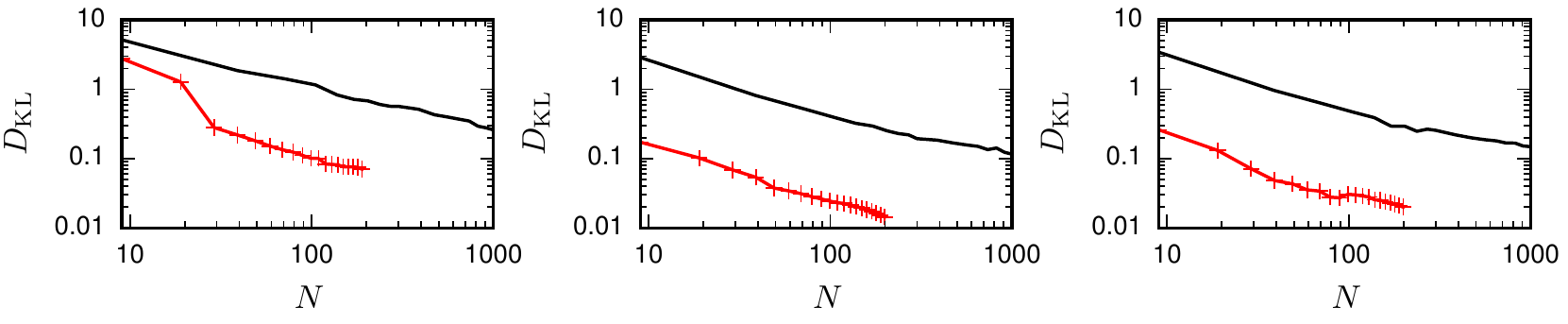}
\caption{Kullback-Leibler divergences as a function of the number of parameters required. Smaller divergences indicate a greater degree of similarity between the two distributions and so a more successful approximation. In black we plot $D_{\rm KL}$ for thinned MCMC chains relative to the long unthinned chain, using kernel density estimates of both and in red we plot values the $D_{\rm KL}$ between the Gaussian mixture approximation and the kernel density estimate of the unthinned chain, with crosses indicating different values of $K$ running from $K=1$ to $K=20$. In the left plot we show values for star A, in the middle star B and star C on the right.
\label{fig:KL}}
\end{figure*}

\begin{figure*}
\includegraphics{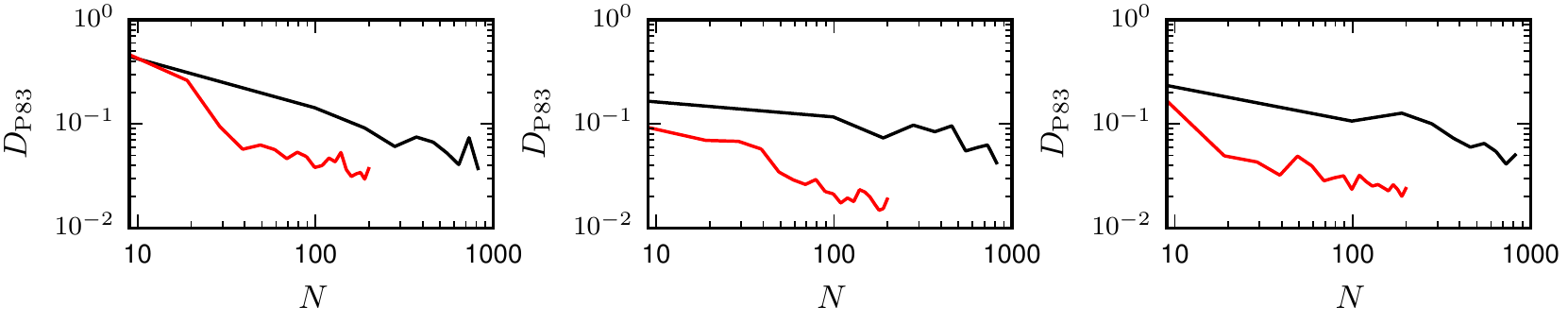}
\caption{\protect\cite{Peacock_only.1983} distances as a function of the number of parameters required. Smaller distances indicate a greater degree of similarity between the two distributions and so a more successful approximation. In black we plot distances for thinned MCMC chains relative to the unthinned chain and in red we plot values the distances between the Gaussian mixture approximation and the unthinned chain. In the left plot we show values for star A, in the middle star B and star C on the right.
\label{fig:D_p83}}
\end{figure*}

One of our primary goals in this paper is to find a compact
description of the marginal likelihood
$\pr(\vy|\mu,a_{4000},R_{5495},\transfermodel,\galaxymodel)$.
It is natural then to ask whether our Gaussian mixture model provides
a more compact summary of this function than, say, a thinned sample of
points from an MCMC chain.
In the following we consider two different measures of how well such
fits reproduce the true marginal likelihood.

\subsubsection{Kullback Leibler divergences}

One way of quantifying the fidelity of different descriptions of the
marginal likelihood is by using the Kullback-Leibler (KL) divergence.
Let $P$ be the true marginalised likelihood and $Q$ a fit from either
the Gaussian mixture or the thinned MCMC sample.
The divergence of the fit $Q$ from the true function~$P$ is given by
\begin{equation}
D_{\rm KL}(P\parallel Q) = \int_{-\infty}^{\infty}\, \d\vtheta \, P(\vtheta) \log \left( \frac{P(\vtheta)}{Q(\vtheta)} \right). \label{eqn:KL}
\end{equation}
This can be recognised as the entropy of $P$ relative to $Q$, a
measure of how much more information there is in $P$ than in the
fitted~$Q$.

One problem with applying this is that we do not know the true
marginal likelihood function~$P(\vtheta)$: we only have discrete samples
$(\vtheta_1,...,\vtheta_N)$ of it from the MCMC chain.
So, to construct our reference $P$ we take a very long chain of
$N\sim10^5$ samples and then, as a simple kernel density estimator, we
replace each sample point $\vtheta_n$ (which has density
$\delta(\vtheta-\vtheta_n)$) by a narrow Gaussian kernel centred on $\vtheta_n$.
Then the value of the function $P$ at any point~$\vtheta$ is given by the
sum of the contributions from all $N\sim10^5$ kernels at that point.

We set the kernel width using ten-fold cross validation.
That is, each point from the MCMC chain is assigned at random to one
of ten subsamples.
Then, for a given trial kernel width, we construct a kernel density
estimate using nine of the ten subsamples.
We use this kernel density estimate to calculate the log-likelihood of
the points in the remaining sub-sample.
This is then repeated for all ten subsamples and the average
log-likelihood found.
By considering a range of kernel widths we can choose an optimum value
by maximising the mean log-likelihood.
Typically the kernel widths found by this procedure are small -- on
the order of 0.01 in $\mu$, for example -- and much smaller than the
bin sizes adopted in Fig.~\ref{fig:marg_ex}.
Consequently, if one applies a binning to the kernel density estimate 
of the marginal likelihood to match that employed in 
Fig.~\ref{fig:marg_ex}, one would obtain a distribution that will 
closely resemble the histograms in Fig.~\ref{fig:marg_ex}.

We use a similar procedure to reconstruct $Q(\vtheta)$ for the thinned
MCMC chains.
We do so by thinning the main MCMC chain and reapplying the
cross-validation procedure to each thinned chain.
Finally we estimate the integral in \eqref{eqn:KL} using Monte Carlo
integration with 10,000 samples drawn from~$P(\vtheta)$.

We are interested in how the KL-divergences from $P(\vtheta)$ of Gaussian
mixture fits and of the thinned MCMC chains scale with the number of
parameters needed to describe each fit.
As discussed in section~\ref{sec:fit} we require $(10 K - 1)$
parameters to describe a $K$ component Gaussian mixture model, whilst
the number of parameters needed to describe a thinned MCMC chain is
the dimensionality (i.e. three) multiplied by the number of samples in
the chain.

In Fig.~\ref{fig:KL} we compare the $D_{\rm KL}$ found using the
Gaussian mixture model approximation to those obtained using thinned
MCMC chains as a function of the number of parameters required.
To achieve a given~$D_{\rm KL}$, the Gaussian mixture model requires
an order of magnitude fewer points than the thinned MCMC chain.

We note that $D_{KL}$ for the Gaussian mixture does not pass below
$\sim 0.02$ for any of the three stars shown.
There are a number of sources of error and noise that will prevent a
perfect agreement between $Q$ and $P$, and so $D_{KL}=0$ being
achieved.
Most fundamentally, the exact marginal likelihood will not, in
general, take a form that can be fit with $K \leq 20$ Gaussian
components.
In particular the marginal likelihood for star A takes a more
complicated form than that of B or C (Fig.~\ref{fig:marg_ex}), which
is reflected in correspondingly large values of $D_{KL}$.
In addition, we do not actually know the exact marginal likelihood.
Instead we have only a noisy kernel-density estimate of it, which
limits our ability to fit smooth functions, such as Gaussian mixtures
to it.
Also, we limit the EM algorithm that fits the Gaussian mixture to a
maximum number of iterations.
Consequently it will generally not achieve the absolute best fit.
The resulting error will be manifested in a small contribution to the
measured $D_{KL}$.
Despite these shortcomings of our $D_{KL}$ tests, we neverthless
believe that it is evident that our Gaussian-mixture fits produce very
good descriptions of the marginal likelihoods.
The $D_{KL} \simeq 0.02$ achieved for stars B and C indicate that our
Gaussian mixture fit differs from the marginal likelihood by, at most,
$\sim 2\%$ on average.

\subsubsection{Kolmogorov--Smirnov tests} 

With one dimensional data it is common to compare samples and/or 
distributions using the  the Kolmogorov-Smirnov distance,
\begin{equation}
D_{\rm KS}(P,Q) \equiv \sup_x |P(x)-Q(x)|, \label{eqn:KS}
\end{equation}
where $P$ and $Q$ are cumulative distributions, either derived
directly from a probability distribution, or found empirically from a
sample of points.
The Kolmogorov-Smirnov distance is not immediately applicable to our
situation, as it is defined only for one dimensional distributions
$P(x)$ and $Q(x)$.
In this one-dimensional case there are only two possible cumulative
distribution functions, either $\pr(x<X)$ or $\pr(x>X)$, each of which
is uniquely defined by the other, because
\begin{equation}
\pr(x \leq X) = 1 - \pr(x >X) .
\label{eqn:cumsum}
\end{equation}

In $p\ge2$ dimensions the notion of a cumulative distribution function
breaks down.
One way of proceeding \citep{Peacock_only.1983} is by constructing
cumulative DFs with respect to the coordinate axes, such as
$\pr(x_1<X_1,x_2<X_2,...,x_p<X_p)$ or
$\pr(x_1>X_1,x_2<X_2,...,x_p>X_p)$ and so on.
For each of our $p=3$ variables we are free to choose either sign of
the inequality when constructing the CDF, giving $2^p=8$ different
possibilities.
We follow \cite{Peacock_only.1983} in calculating the one-dimensional
KS distance for all 8 possibilities of CDFs for $P$ and~$Q$, then
taking the maximum such distance as our measure of the ``similarity''
of the two functions.
\footnote {The multidimensional analogue of
  equation~\eqref{eqn:cumsum} means that any one of these CDFs is
  completely determined by the other $2^p-1$.
  So, there are $2^p-1=7$ independent CDFs, but this redundancy does
  not affect Peacock's argument. }

An advantage of this scheme over the KL divergence is that it can be
applied directly to the samples from MCMC chains: it avoids the need
for kernel density estimates of either $P$ or, in the case of thinned
MCMC chains, $Q$.
Fig.~\ref{fig:D_p83} shows the results.
As with the KL-divergences, for a given number of parameters the
Gaussian mixture model provides a far better approximation to the
marginal likelihood than a thinned MCMC chain can.

\section{Summary}\label{sec:summary}

We have considered how one should measure the distance and extinction
to individual stars for use in constructing extinction maps of the
whole galaxy.
We advocate the use of monochromatic extinctions, since, unlike band
pass measures such as $A_V$ and $E(B-V)$, monochromatic extinctions
are linear functions of the dust column density and are
independent of the source SED.
In particular we suggest the use of $A_{4000}$, the monochromatic
extinction at $4000$~\AA \, because of its insensitivity to the dust
grain size distribution.

We have developed one way of calculating the marginal likelihood 
$\pr(\tilde{\vy} |\mu, a_{4000}, R_{5495}, \transfermodel, 
\galaxymodel)$ by marginalising the (unknown and, for our purposes, 
uninteresting) fundamental parameters of the star in order to estimate
the marginal likelihood.
As this integration is not possible analytically, we suggest a scheme
for doing so using MCMC methods, specifically the affine invariant
ensemble sampler of \cite{Goodman_Weare.2010}

We find that the resulting marginal likelihood function can be
described very well using a Gaussian mixture model composed of only
$K\simeq5$ Gaussians.
Using thinned MCMC chains would require vastly more parameters to
achieve the same level of fidelity.
Having such a compact description of $\pr(\tilde{\vy} |\mu, a_{4000},
R_{5495} \transfermodel, \galaxymodel)$ is vital when one is
constructing maps from large catalogues of stars.
Another advantage of expressing the marginal likelihood as a sum of
Gaussians is that it makes further marginalisation of any or all of
the parameters $(\mu,a_{4000},R_{5495})$ straightforward.
This is particularly important if one models the dust density
distribution as a Gaussian random field \citep{Sale_Magorrian.2014}.

In common with \cite{Green_Schlafly.2014}, the approach adopted in
\cite{Sale_Magorrian.2014} is to split the production of
three-dimensional dust maps into two distinct steps.
First we estimate the marginal likelihood $\pr(\tilde{\vy} |\mu,
a_{4000}, R_{5495}, \transfermodel, \galaxymodel)$ of distance
modulus~$\mu$ and (log) extinction $a_{4000}$ to each star in the
catalogue.
Then we construct maps from these distances and extinctions.
The method we present in this paper for carrrying out the first of
these two steps is very similar to the method
\cite{Green_Schlafly.2014} use for calculating their posterior pdf
$\pr(\mu,A|\tilde\vy)$.
The most important differences are that we use monochromatic
extinctions and we return the result in a compact multi-Gaussian form.
Our sample of Galactic plane stars meant that we could reasonably use
a simple prior~$\galaxymodel$ on stellar distances and intrinsic
parameters~$\vx$: this is easy to change for more extended samples.

The alternative to these two-step approaches would be to infer
simultaneously the distance--extinction relationship and the
properties of all the stars that trace it \citep{Sale_only.2012}.
The benefit of this is that MCMC schemes operating in the extended
space of the stars' intrinsic parameters and their $(\mu, a_{4000},
R_{5495})$ would tend to avoid regions of $(\mu, a_{4000}, R_{5495})$
that are a posteriori unlikely, reducing the computing load.
The downside is that parallelization becomes very difficult, making it
infeasible to scale up to large datasets.
In contrast, in the two-step procedure one has no way of knowing what
portions of $(\mu, a_{4000}, R_{5495})$ parameter space are going to
be important, and so it has to be explored thoroughly.
But this is a small price to pay for the trivial parallelization
opportunities.

The software libraries used to obtain the results in this paper are
available online, including a library for manipulating
isochrones\footnote{\url{https://github.com/stuartsale/iso_lib}} and
the code used to sample the marginal likelihood and fit it with a
Gaussian mixture
model\footnote{\url{https://github.com/stuartsale/marg_iso}}.

\section*{Acknowledgements}

The research leading to the results presented here was supported by
the United Kingdom Science Technology and Facilities Council (STFC,
ST/K00106X/1), the European Research Council under the European
Union’s Seventh Framework Programme (FP7/2007-2013)/ERC grant
agreement no. 321067.
JM thanks the Institut d'Astrophysique de Paris for their hospitality
and Ville de Paris for support through their ``Research in Paris''
programme.

\bibliography{astroph_3,bibliography-2}

\begin{thebibliography}{54}
\expandafter\ifx\csname natexlab\endcsname\relax\def\natexlab#1{#1}\fi

\bibitem[{{Bailer-Jones}(2011)}]{Bailer-Jones_only.2011}
{Bailer-Jones} C.~A.~L., 2011, \mnras, 411, 435

\bibitem[{{Barentsen} {et~al}\mbox{.}(2014){Barentsen}, {Farnhill}, {Drew},
  {Gonz{\'a}lez-Solares}, {Greimel}, {Irwin}, {Miszalski}, {Ruhland}, {Groot},
  {Mampaso}, {Sale}, {Henden}, {Aungwerojwit}, {Barlow}, {Carter}, {Corradi},
  {Drake}, {Eisl{\"o}ffel}, {Fabregat}, {G{\"a}nsicke}, {Gentile Fusillo},
  {Greiss}, {Hales}, {Hodgkin}, {Huckvale}, {Irwin}, {King}, {Knigge},
  {Kupfer}, {Lagadec}, {Lennon}, {Lewis}, {Mohr-Smith}, {Morris}, {Naylor},
  {Parker}, {Phillipps}, {Pyrzas}, {Raddi}, {Roelofs}, {Rodr{\'{\i}}guez-Gil},
  {Sabin}, {Scaringi}, {Steeghs}, {Suso}, {Tata}, {Unruh}, {van Roestel},
  {Viironen}, {Vink}, {Walton}, {Wright}, \&
  {Zijlstra}}]{Barentsen_Farnhill.2014}
{Barentsen} G. {et~al.}, 2014, \mnras, 444, 3230

\bibitem[{{Berry} {et~al}\mbox{.}(2012){Berry}, {Ivezi{\'c}}, {Sesar},
  {Juri{\'c}}, {Schlafly}, {Bellovary}, {Finkbeiner}, {Vrbanec}, {Beers},
  {Brooks}, {Schneider}, {Gibson}, {Kimball}, {Jones}, {Yoachim}, {Krughoff},
  {Connolly}, {Loebman}, {Bond}, {Schlegel}, {Dalcanton}, {Yanny}, {Majewski},
  {Knapp}, {Gunn}, {Allyn Smith}, {Fukugita}, {Kent}, {Barentine},
  {Krzesinski}, \& {Long}}]{Berry_Ivezic.2012}
{Berry} M. {et~al.}, 2012, \apj, 757, 166

\bibitem[{{Bessell} {et~al}\mbox{.}(2011){Bessell}, {Bloxham}, {Schmidt},
  {Keller}, {Tisserand}, \& {Francis}}]{Bessell_Bloxham.2011}
{Bessell} M., {Bloxham} G., {Schmidt} B., {Keller} S., {Tisserand} P.,
  {Francis} P., 2011, \pasp, 123, 789

\bibitem[{{Bessell}(1990)}]{Bessell_only.1990}
{Bessell} M.~S., 1990, \pasp, 102, 1181

\bibitem[{{Bessell}(2005)}]{Bessell_only.2005}
---, 2005, \araa, 43, 293

\bibitem[{{Bressan} {et~al}\mbox{.}(2012){Bressan}, {Marigo}, {Girardi},
  {Salasnich}, {Dal Cero}, {Rubele}, \& {Nanni}}]{Bressan_Marigo.2012}
{Bressan} A., {Marigo} P., {Girardi} L., {Salasnich} B., {Dal Cero} C.,
  {Rubele} S., {Nanni} A., 2012, \mnras, 427, 127

\bibitem[{{Cardelli}, {Clayton} \& {Mathis}(1989){Cardelli}, {Clayton}, \&
  {Mathis}}]{Cardelli_Clayton.1989}
{Cardelli} J.~A., {Clayton} G.~C., {Mathis} J.~S., 1989, \apj, 345, 245

\bibitem[{{Carrasco Kind} \& {Brunner}(2014)}]{CarrascoKind_Brunner.2014}
{Carrasco Kind} M., {Brunner} R.~J., 2014, \mnras, 441, 3550

\bibitem[{{Casagrande} \& {VandenBerg}(2014)}]{Casagrande_VandenBerg.2014}
{Casagrande} L., {VandenBerg} D.~A., 2014, \mnras, 444, 392

\bibitem[{{Castelli} \& {Kurucz}(2003)}]{Castelli_Kurucz.2003}
{Castelli} F., {Kurucz} R.~L., 2003, in IAU Symposium, {Piskunov} N., {Weiss}
  W.~W., {Gray} D.~F., eds., pp. 20P--+

\bibitem[{{Cohen}, {Wheaton} \& {Megeath}(2003){Cohen}, {Wheaton}, \&
  {Megeath}}]{Cohen_Wheaton.2003}
{Cohen} M., {Wheaton} W.~A., {Megeath} S.~T., 2003, \aj, 126, 1090

\bibitem[{{Crawford} \& {Barnes}(1970)}]{Crawford_Barnes.1970}
{Crawford} D.~L., {Barnes} J.~V., 1970, \aj, 75, 978

\bibitem[{{Crawford} \& {Mander}(1966)}]{Crawford_Mander.1966}
{Crawford} D.~L., {Mander} J., 1966, \aj, 71, 114

\bibitem[{{di Francesco} {et~al}\mbox{.}(2010){di Francesco}, {Sadavoy},
  {Motte}, {Schneider}, {Hennemann}, {Csengeri}, {Bontemps}, {Balog},
  {Zavagno}, {Andr{\'e}}, {Saraceno}, {Griffin}, {Men'shchikov}, {Abergel},
  {Baluteau}, {Bernard}, {Cox}, {Deharveng}, {Didelon}, {di Giorgio},
  {Hargrave}, {Huang}, {Kirk}, {Leeks}, {Li}, {Marston}, {Martin}, {Minier},
  {Molinari}, {Olofsson}, {Persi}, {Pezzuto}, {Russeil}, {Sauvage},
  {Sibthorpe}, {Spinoglio}, {Testi}, {Teyssier}, {Vavrek}, {Ward-Thompson},
  {White}, {Wilson}, \& {Woodcraft}}]{DiFrancesco_Sadavoy.2010}
{di Francesco} J. {et~al.}, 2010, \aap, 518, L91

\bibitem[{{Draine}(2003)}]{Draine_only.2003}
{Draine} B.~T., 2003, \araa, 41, 241

\bibitem[{{Drew} {et~al}\mbox{.}(2014){Drew}, {Gonzalez-Solares}, {Greimel},
  {Irwin}, {K{\"u}pc{\"u} Yoldas}, {Lewis}, {Barentsen}, {Eisl{\"o}ffel},
  {Farnhill}, {Martin}, {Walsh}, {Walton}, {Mohr-Smith}, {Raddi}, {Sale},
  {Wright}, {Groot}, {Barlow}, {Corradi}, {Drake}, {Fabregat}, {Frew},
  {G{\"a}nsicke}, {Knigge}, {Mampaso}, {Morris}, {Naylor}, {Parker},
  {Phillipps}, {Ruhland}, {Steeghs}, {Unruh}, {Vink}, {Wesson}, \&
  {Zijlstra}}]{Drew_Gonzalez-Solares.2014}
{Drew} J.~E. {et~al.}, 2014, \mnras, 440, 2036

\bibitem[{{Drew} {et~al}\mbox{.}(2005){Drew}, {Greimel}, {Irwin},
  {Aungwerojwit}, {Barlow}, {Corradi}, {Drake}, {G{\"a}nsicke},
  {et~al.}}]{Drew_Greimel.2005}
---, 2005, \mnras, 362, 753

\bibitem[{{Fitzpatrick}(2004)}]{Fitzpatrick_only.2004}
{Fitzpatrick} E.~L., 2004, in Astronomical Society of the Pacific Conference
  Series, Vol. 309, Astrophysics of Dust, {Witt} A.~N., {Clayton} G.~C.,
  {Draine} B.~T., eds., p.~33

\bibitem[{{Fitzpatrick} \& {Massa}(2007)}]{Fitzpatrick_Massa.2007}
{Fitzpatrick} E.~L., {Massa} D., 2007, \apj, 663, 320

\bibitem[{{Foreman-Mackey} {et~al}\mbox{.}(2013){Foreman-Mackey}, {Hogg},
  {Lang}, \& {Goodman}}]{Foreman-Mackey_Hogg.2013}
{Foreman-Mackey} D., {Hogg} D.~W., {Lang} D., {Goodman} J., 2013, \pasp, 125,
  306

\bibitem[{{Golay}(1974)}]{Golay_only.1974}
{Golay} M., ed., 1974, Astrophysics and Space Science Library, Vol.~41,
  {Introduction to astronomical photometry}

\bibitem[{Goodman \& Weare(2010)}]{Goodman_Weare.2010}
Goodman J., Weare J., 2010, Communications in Applied Mathematics and
  Computational Science, 5, 65

\bibitem[{{Green} {et~al}\mbox{.}(2014){Green}, {Schlafly}, {Finkbeiner},
  {Juri{\'c}}, {Rix}, {Burgett}, {Chambers}, {Draper}, {Flewelling},
  {Kudritzki}, {Magnier}, {Martin}, {Metcalfe}, {Tonry}, {Wainscoat}, \&
  {Waters}}]{Green_Schlafly.2014}
{Green} G.~M. {et~al.}, 2014, \apj, 783, 114

\bibitem[{{Hanson} \& {Bailer-Jones}(2014)}]{Hanson_Bailer-Jones.2014}
{Hanson} R.~J., {Bailer-Jones} C.~A.~L., 2014, \mnras, 438, 2938

\bibitem[{{Hewett} {et~al}\mbox{.}(2006){Hewett}, {Warren}, {Leggett}, \&
  {Hodgkin}}]{Hewett_Warren.2006}
{Hewett} P.~C., {Warren} S.~J., {Leggett} S.~K., {Hodgkin} S.~T., 2006, \mnras,
  367, 454

\bibitem[{{Hou} {et~al}\mbox{.}(2012){Hou}, {Goodman}, {Hogg}, {Weare}, \&
  {Schwab}}]{Hou_Goodman.2012}
{Hou} F., {Goodman} J., {Hogg} D.~W., {Weare} J., {Schwab} C., 2012, \apj, 745,
  198

\bibitem[{{Husser} {et~al}\mbox{.}(2013){Husser}, {Wende-von Berg}, {Dreizler},
  {Homeier}, {Reiners}, {Barman}, \& {Hauschildt}}]{Husser_Wende-vonBerg.2013}
{Husser} T.-O., {Wende-von Berg} S., {Dreizler} S., {Homeier} D., {Reiners} A.,
  {Barman} T., {Hauschildt} P.~H., 2013, \aap, 553, A6

\bibitem[{{Lallement} {et~al}\mbox{.}(2014){Lallement}, {Vergely}, {Valette},
  {Puspitarini}, {Eyer}, \& {Casagrande}}]{Lallement_Vergely.2014}
{Lallement} R., {Vergely} J.-L., {Valette} B., {Puspitarini} L., {Eyer} L.,
  {Casagrande} L., 2014, \aap, 561, A91

\bibitem[{{Lucas} {et~al}\mbox{.}(2008){Lucas}, {Hoare}, {Longmore},
  {Schr{\"o}der}, {Davis}, {Adamson}, {Bandyopadhyay}, {de Grijs},
  {et~al.}}]{Lucas_Hoare.2008}
{Lucas} P.~W. {et~al.}, 2008, \mnras, 391, 136

\bibitem[{{Luck} \& {Lambert}(2011)}]{Luck_Lambert.2011}
{Luck} R.~E., {Lambert} D.~L., 2011, \aj, 142, 136

\bibitem[{{Magorrian}(2014)}]{Magorrian_only.2014}
{Magorrian} J., 2014, \mnras, 437, 2230

\bibitem[{{Maiz Apell{\'a}niz}(2013)}]{MaizApellaniz_only.2013}
{Maiz Apell{\'a}niz} J., 2013, in Highlights of Spanish Astrophysics
  VII, {Guirado} J.~C., {Lara} L.~M., {Quilis} V., {Gorgas} J., eds., pp.
  583--589

\bibitem[{{Majewski}, {Zasowski} \& {Nidever}(2011){Majewski}, {Zasowski}, \&
  {Nidever}}]{Majewski_Zasowski.2011}
{Majewski} S.~R., {Zasowski} G., {Nidever} D.~L., 2011, \apj, 739, 25

\bibitem[{{Marshall} {et~al}\mbox{.}(2006){Marshall}, {Robin}, {Reyl{\'e}},
  {Schultheis}, \& {Picaud}}]{Marshall_Robin.2006}
{Marshall} D.~J., {Robin} A.~C., {Reyl{\'e}} C., {Schultheis} M., {Picaud} S.,
  2006, \aap, 453, 635

\bibitem[{{McCall}(2004)}]{McCall_only.2004}
{McCall} M.~L., 2004, \aj, 128, 2144

\bibitem[{{Munari} {et~al}\mbox{.}(2005){Munari}, {Sordo}, {Castelli}, \&
  {Zwitter}}]{Munari_Sordo.2005}
{Munari} U., {Sordo} R., {Castelli} F., {Zwitter} T., 2005, \aap, 442, 1127

\bibitem[{{O'Donnell}(1994)}]{ODonnell_only.1994}
{O'Donnell} J.~E., 1994, \apj, 422, 158

\bibitem[{{Patat} {et~al}\mbox{.}(2011){Patat}, {Moehler}, {O'Brien}, {Pompei},
  {Bensby}, {Carraro}, {de Ugarte Postigo}, {Fox}, {Gavignaud}, {James},
  {Korhonen}, {Ledoux}, {Randall}, {Sana}, {Smoker}, {Stefl}, \&
  {Szeifert}}]{Patat_Moehler.2011}
{Patat} F. {et~al.}, 2011, \aap, 527, A91

\bibitem[{{Peacock}(1983)}]{Peacock_only.1983}
{Peacock} J.~A., 1983, \mnras, 202, 615

\bibitem[{Pedregosa {et~al}\mbox{.}(2011)Pedregosa, Varoquaux, Gramfort,
  Michel, Thirion, Grisel, Blondel, Prettenhofer, Weiss, Dubourg, Vanderplas,
  Passos, Cournapeau, Brucher, Perrot, \& Duchesnay}]{Pedregosa_Varoquaux.2011}
Pedregosa F. {et~al.}, 2011, Journal of Machine Learning Research, 12, 2825

\bibitem[{{Planck Collaboration} {et~al}\mbox{.}(2014){Planck Collaboration},
  {Abergel}, {Ade}, {Aghanim}, {Alves}, {Aniano}, {Armitage-Caplan}, {Arnaud},
  {Ashdown}, {Atrio-Barandela}, \& et~al.}]{PlanckCollaboration_Abergel.2014a}
{Planck Collaboration} {et~al.}, 2014, \aap, 571, A11

\bibitem[{{Sale}(2012)}]{Sale_only.2012}
{Sale} S.~E., 2012, \mnras, 427, 2119

\bibitem[{{Sale} {et~al}\mbox{.}(2014){Sale}, {Drew}, {Barentsen}, {Farnhill},
  {Raddi}, {Barlow}, {Eisl{\"o}ffel}, {Vink}, {Rodr{\'{\i}}guez-Gil}, \&
  {Wright}}]{Sale_Drew.2014}
{Sale} S.~E. {et~al.}, 2014, \mnras, 443, 2907

\bibitem[{{Sale} {et~al}\mbox{.}(2009){Sale}, {Drew}, {Unruh}, {Irwin},
  {Knigge}, {Phillipps}, {Zijlstra}, {G{\"a}nsicke}, {Greimel}, {Groot},
  {Mampaso}, {Morris}, {Napiwotzki}, {Steeghs}, \& {Walton}}]{Sale_Drew.2009}
---, 2009, \mnras, 392, 497

\bibitem[{{Sale} \& {Magorrian}(2014)}]{Sale_Magorrian.2014}
{Sale} S.~E., {Magorrian} J., 2014, \mnras, 445, 256

\bibitem[{{Schlafly} \& {Finkbeiner}(2011)}]{Schlafly_Finkbeiner.2011}
{Schlafly} E.~F., {Finkbeiner} D.~P., 2011, \apj, 737, 103

\bibitem[{{Schlegel}, {Finkbeiner} \& {Davis}(1998){Schlegel}, {Finkbeiner}, \&
  {Davis}}]{Schlegel_Finkbeiner.1998}
{Schlegel} D.~J., {Finkbeiner} D.~P., {Davis} M., 1998, \apj, 500, 525

\bibitem[{Schwarz {et~al}\mbox{.}(1978)Schwarz {et~al.}}]{Schwarz_only.1978}
Schwarz G., {et~al.}, 1978, The annals of statistics, 6, 461

\bibitem[{{Stead} \& {Hoare}(2009)}]{Stead_Hoare.2009}
{Stead} J.~J., {Hoare} M.~G., 2009, \mnras, 400, 731

\bibitem[{{Straizys} \& {Kuriliene}(1981)}]{Straizys_Kuriliene.1981}
{Straizys} V., {Kuriliene} G., 1981, \apss, 80, 353

\bibitem[{{Stubbs} {et~al}\mbox{.}(2010){Stubbs}, {Doherty}, {Cramer},
  {Narayan}, {Brown}, {Lykke}, {Woodward}, \& {Tonry}}]{Stubbs_Doherty.2010}
{Stubbs} C.~W., {Doherty} P., {Cramer} C., {Narayan} G., {Brown} Y.~J., {Lykke}
  K.~R., {Woodward} J.~T., {Tonry} J.~L., 2010, \apjs, 191, 376

\bibitem[{{Vergely} {et~al}\mbox{.}(2001){Vergely}, {Freire Ferrero},
  {Siebert}, \& {Valette}}]{Vergely_FreireFerrero.2001}
{Vergely} J.-L., {Freire Ferrero} R., {Siebert} A., {Valette} B., 2001, \aap,
  366, 1016

\bibitem[{{Weingartner} \& {Draine}(2001)}]{Weingartner_Draine.2001}
{Weingartner} J.~C., {Draine} B.~T., 2001, \apj, 548, 296

\end{thebibliography}

\appendix

\section{Tabulated response to $A_{4000}$ for a variety of photometric bands} \label{app:abs}

\begin{table*}
\begin{tabular}{c|c|p{10cm}}
System & Filters & Source \\
\hline
Bessell & $UBVRI$ & \cite{Bessell_only.1990}\\
Str\"omgren & $ubvy\Hbeta_{\rm narrow}\Hbeta_{wide}$ & \cite{Crawford_Barnes.1970}, \cite{Crawford_Mander.1966} \\
\hline
2MASS & $JHK_s$ & \cite{Cohen_Wheaton.2003} \\
Gaia & $G$ & \url{http://www.cosmos.esa.int/web/gaia/transmissionwithoriginal} \\
INT (IPHAS/UVEX) & $Ugri\Halpha$ & \url{http://www.ing.iac.es/astronomy/instruments/wfc/} \\
PAN-STARRS & $grizy$ & \cite{Stubbs_Doherty.2010} \\
SDSS & $ugriz$ & \url{https://www.sdss3.org/instruments/camera.php#Filters} \\
Skymapper & $uvgriz$ & \cite{Bessell_Bloxham.2011} \\
UKIDSS & $ZYJHK$ & \cite{Hewett_Warren.2006} \\
VISTA & $ZYJHK_s$& \url{http://www.eso.org/sci/facilities/paranal/instruments/vircam/inst.html} \\
VST & $ugri\Halpha$ & \url{http://www.eso.org/sci/facilities/paranal/instruments/omegacam/tools.html}, \cite{Drew_Gonzalez-Solares.2014} \\
\end{tabular}
\caption{A list of the systems and filters for which we tabulate the response to extinction.
\label{tab:filters} }
\end{table*}

We make available with this paper\footnote{\url{https://github.com/stuartsale/A4000_coeffs} }
a tabulation of the the coefficients $\extnlin_X$ and $\extnquad_X$ 
of~\eqref{eqn:AXalphabeta} for a variety of filters and the full range
of \cite{Fitzpatrick_only.2004} reddening laws.
We do this for SEDs along a solar metallicity main sequence, defined 
in $(T_{\rm eff}, \log g)$ by \cite{Straizys_Kuriliene.1981}.
We also include a Rayleigh-Jeans spectrum, to demonstrate the limiting 
behaviour for extremely hot stars.

In table~\ref{tab:filters} we list the photometric systems and their 
constituent filters that we employ.
For all the survey filter sets we also employ the detector 
quantum efficiency curve and atmospheric transmission for the 
instrument and site used.

We include two `standard' filter sets: the \cite{Bessell_only.1990} 
$UBVRI$ set and a Str\"omgren filter set with the $uvby$ transmissions
taken from \cite{Crawford_Barnes.1970} and $\Hbeta_{\rm wide}$ and 
$\Hbeta_{\rm narrow}$ from \cite{Crawford_Mander.1966}.
We present results for these filters using the INT/WFC CCD quantum 
efficiency and the \cite{Patat_Moehler.2011} model for atmospheric 
absorption at Cerro Paranal.
In addition, for reference purposes, we also provide results for the 
\cite{Bessell_only.1990} filter set with no atmospheric absorption and
a 100\% efficient detector.

A sample of the table of values of $\extnlin_X$ and $\extnquad_X$ is given 
in table~\ref{tab:sample}.

\begin{table*}
\tiny
\begin{tabular}{*{4}{c}|*{6}{c}|*{6}{c} }
\multirow{2}{*}{$R_{5495}$} & \multirow{2}{*}{Spectral Type} & \multirow{2}{*}{$T_{\rm eff}$} & \multirow{2}{*}{$\log g$} & \multicolumn{6}{| c |}{2MASS} & \multicolumn{6}{| c |}{Bessell} \\ 
 & & &  & $\extnlin_{ J }$ & $\extnquad_{ J }$ & $\extnlin_{ H }$ & $\extnquad_{ H }$ & $\extnlin_{ Ks }$ & $\extnquad_{ Ks }$ & $\extnlin_{ U }$ & $\extnquad_{ U }$ & $\extnlin_{ B }$ & $\extnquad_{ B }$ & $\extnlin_{ V }$ & $\extnquad_{ V }$ \\ 
\hline 
2.1 & Rayleigh-Jeans & -- & -- & 0.12046 & -8.688E-05 & 0.07181 & -1.816E-05 & 0.04307 & -4.887E-06 & 1.154 & -0.003527 & 0.8794 & -0.005505 & 0.5961 & -0.002175 \\ 
2.1 & O9 & 33500 & 3.95 & -- & -- & -- & -- & -- & -- & 1.15 & -0.003517 & 0.8754 & -0.005474 & 0.5953 & -0.002184 \\ 
2.1 & B0 & 31480 & 4.00 & -- & -- & -- & -- & -- & -- & 1.149 & -0.003501 & 0.8751 & -0.005468 & 0.5951 & -0.002181 \\ 
2.1 & B1 & 26490 & 4.00 & -- & -- & -- & -- & -- & -- & 1.146 & -0.003462 & 0.8737 & -0.005446 & 0.5946 & -0.002181 \\ 
2.1 & B2 & 23010 & 4.06 & -- & -- & -- & -- & -- & -- & 1.143 & -0.003437 & 0.8727 & -0.005422 & 0.5944 & -0.002182 \\ 
2.1 & B3 & 19320 & 4.06 & -- & -- & -- & -- & -- & -- & 1.137 & -0.003378 & 0.871 & -0.005389 & 0.5938 & -0.002185 \\ 
2.1 & B5 & 15420 & 4.10 & -- & -- & -- & -- & -- & -- & 1.13 & -0.003264 & 0.8687 & -0.005348 & 0.5932 & -0.002189 \\ 
2.1 & B6 & 14190 & 4.09 & -- & -- & -- & -- & -- & -- & 1.127 & -0.003211 & 0.868 & -0.005336 & 0.593 & -0.002189 \\ 
2.1 & B7 & 12790 & 4.07 & -- & -- & -- & -- & -- & -- & 1.118 & -0.003045 & 0.8662 & -0.005303 & 0.5926 & -0.00219 \\ 
2.1 & B8 & 11510 & 4.07 & 0.12015 & -8.678E-05 & 0.07162 & -1.814E-05 & 0.04305 & -4.899E-06 & 1.116 & -0.003041 & 0.8651 & -0.00528 & 0.5924 & -0.00219 \\ 
\end{tabular}

\caption{An extract from the compilation of the coefficients $\extnlin_X$ and $\extnquad_X$ of~\eqref{eqn:AXalphabeta} for the filters listed in table~\ref{tab:filters} and the \protect\cite{Fitzpatrick_only.2004} reddening laws.
\label{tab:sample} }
\end{table*}

\end{document}